\documentclass[fleqn,usenatbib]{mnras}

\usepackage{newtxtext,newtxmath}

\usepackage[T1]{fontenc}
\usepackage{ae,aecompl}
\usepackage{xcolor}

\usepackage{graphicx}	
\usepackage{amsmath}	
\usepackage{scalerel}

\usepackage[normalem]{ulem}

\newcommand{\pd}[1]{\frac{\partial}{\partial {#1}}}
\newcommand{\D}{\mathrm{d}}

\newcommand{\abs}[1]{\left|#1\right|}

\title[BZ jets in galaxy formation simulations]{Blandford-Znajek jets in galaxy formation simulations: method and implementation}

\author[R. Y. Talbot et al.]{
Rosie Y. Talbot$^{1}$\thanks{E-mail: rt421@cam.ac.uk (RYT)},
Martin A. Bourne$^{1,2}$ and Debora Sijacki$^{1,2}$
\\
$^{1}$Institute of Astronomy, University of Cambridge, Madingley Road, Cambridge, CB3 0HA, UK\\
$^{2}$Kavli Institute for Cosmology, University of Cambridge, Madingley Road, Cambridge, CB3 0HA, UK
}

\date{Submitted to MNRAS}
\pubyear{2020}
\begin{document}
\label{firstpage}
\pagerange{\pageref{firstpage}--\pageref{lastpage}}
\maketitle

\begin{abstract}
Jets launched by active galactic nuclei (AGN) are believed to play a significant role in shaping the properties of galaxies and provide an energetically viable mechanism through which galaxies can become quenched. Here we present a novel AGN feedback model, which we have incorporated into the \textsc{arepo} code, that evolves the black hole mass and spin as the accretion flow proceeds through a thin $\alpha$-disc which we self-consistently couple to a Blandford-Znajek jet. We apply our model to the central region of a typical radio-loud Seyfert galaxy embedded in a hot circumgalactic medium (CGM). We find that jets launched into high pressure environments thermalise efficiently due to the formation of recollimation shocks and the vigorous instabilities that these shocks excite increase the efficiency of the mixing of CGM and jet material. The beams of more overpressured jets, however, are not as readily disrupted by instabilities so the majority of the momentum flux at the jet base is retained out to the head, where the jet terminates in a reverse shock. All jets entrain a significant amount of cold circumnuclear disc material which, while energetically insignificant, dominates the lobe mass together with the hot, entrained CGM material. The jet power evolves significantly due to effective self-regulation by the black hole, fed by secularly-driven, intermittent mass flows. The direction of jets launched directly into the circumnuclear disc changes considerably due to effective Bardeen-Petterson torquing. Interestingly, these jets obliterate the innermost regions of the disc and drive large-scale, multi-phase, turbulent, bipolar outflows.
\end{abstract}




\section{Introduction}
In the centres of many galaxy clusters the hot, X-ray emitting gas is observed to have rapid ($\lesssim1\; {\rm Gyr}$) cooling times. Without invoking some heating mechanism, significant quantities of cold gas should flow towards the centre of these clusters \citep{2006PetersonFabian, 1994Fabian}. Many such clusters, however, do not show evidence of high levels of star formation \citep{2016Cooke+,2010Donahue+} nor do they have large reservoirs of molecular gas \citep{2019Fogarty+, 2019Russel+, 2017Russel+}, both of which should be present if such a cooling flow were to exist. It naturally follows that some sort of heating process must be at play, offsetting radiative cooling and ensuring that the intracluster medium (ICM) remains in approximate thermal equilibrium. Jet feedback from the AGN associated with the central brightest cluster galaxy is often invoked as such a mechanism \citep{2012McNamaraNulsen, 2007McNamaraNulsen} and in many cases there is direct observational evidence of the interaction of jet lobes with the surrounding medium with estimated cavity energies comparable to that required to offset catastrophic cooling. 

AGN jet feedback, however, is not confined just to these most massive of systems. Indeed, there is growing evidence, both observational and theoretical, that AGN feedback is acting in a variety of galaxy contexts, ranging from spirals \citep[e.g.][]{2015Mao+, 2008Croston+,2001Ledlow+} to dwarfs \citep[e.g.][]{2019Koudmani+, 2019Mezcua+, 2018Mezcua+, 2018Chilingarian+, 2018Penny+, 2017Silk}. Galaxy formation simulations, however, typically disfavour strong AGN feedback in low mass galaxies and, specifically, the role of AGN jets in dwarfs and spirals is largely unexplored. 

There is clear observational evidence for radio jets in Seyfert galaxies \citep{2017Williams+,2012Mingo+,2011Mingo+,2006Gallimore+,2006HotaSaikia, 1999Morganti+}. Interestingly, some studies find little evidence for correlation between the jet axis and the major axis of the host galaxy in these systems \citep{1999NagarWilson, 1998Clarke+,1997Schmitt+} and some show signs of `bent' radio jets which could be due to jet precession or interaction with their surroundings. The lack of correlation could come about due to the large-scale gas angular momentum not being aligned with that of the innermost regions of the accretion disc \citep{2000Kinney+,1997Pringle,1996Pringle}. Upcoming new observational data from Lynx, Athena and SKA will shed new light on these issues by allowing us to investigate these energetically moderate sources in much greater detail both locally and at high redshifts.

There are clearly numerous open questions surrounding the nature of radio-mode feedback: `Can the energy from the jets be efficiently and (largely) isotropically communicated to the surrounding medium and if so, what are the dominant physical mechanisms that are responsible?', `How are these jets launched and what is their duty cycle?', `What processes drive the transition to quasar-mode feedback?' to name but a few.

A large body of work studying X-ray black hole binaries has provided crucial insights as to the transition between low and high accretion state \citep{2014BegelmanArmitage, 2004Fender+}. Theoretical models of AGN accretion and feedback that aim to motivate the transition from radio to quasar mode feedback have been developed in analogy with this \citep[see e.g.][]{2014YuanNarayan, 2008MerloniHeinz, 2005Churazov+}. Direct observational evidence, however, is much less conclusive due to the significantly longer timescales associated with supermassive black hole accretion relative to those of X-ray binary systems.

Simulations of AGN jets prove a vital tool to shed further light on these questions. The processes expected to be at play, however, span a vast range of scales and, hence, many different `flavours' of simulations have developed, each of which is specialised to probe the relevant physics in different regimes. 

On the smallest scales, general relativistic magnetohydrodynamic (GRMHD) simulations have proved an effective tool to investigate the formation of these jets and how they interact with the accretion flow that feeds the black hole \citep{2019Chatterjee+, 2018Liska+, 2015SadowskiNarayan,2011Tchekhovskoy+,2010Penna+,2006McKinney}. In these simulations, crucially, the jet launching scale is resolvable and some simulations find that evacuated funnels of outflowing magnetically dominated plasma develop self-consistently along the black hole spin axis. Due to computational constraints, these simulations mainly focus on thick discs which are expected to form when the accretion rate is significantly sub or super-Eddington. It is only very recently, thanks to significant computational advances, that the thin-disc paradigm has been started to be explored. For example, \citet{2020Liska+,2019Liska+} have shown that the inner regions of initially misaligned, thin discs are torqued into alignment with the equatorial plane of the black hole, leading to the formation of warped discs (in accordance with the predictions of \cite{1975BardeenPetterson}). The fact that these thin discs are able to launch precessing, relativistic, magnetised jets raises the question of whether thick discs are a necessary requirement for the launching of radio jets.

There is a significant body of work that examines AGN jet launching on galaxy scales. These works investigate how the jet interacts with a multiphase interstellar medium (ISM) and how the gas structure in the vicinity of the black hole affects the resulting jet morphology on large scales. They also play a vital role in understanding the conditions under which jet launching may lead to enhanced star formation rates and those under which we would expect star formation to be quenched \citep{2016Wagner+}. To this end, some works consider the interaction of the jet with a single cold cloud \citep{2008Antonuccio-DeloguSilk}, others model a turbulent two-phase ISM \citep{2017Asahina+, 2016Mukherjee+,2011WagnerBicknell} which may be confined to a disc \citep{2018Cielo+2,2018Mukherjee+, 2012Gaibler+,2011Gaibler+}. 

Moving up to larger scales, much work has been done on jet propagation hydrostatic atmospheres and in idealised cluster setups, with the aim of better understanding the jet-ICM interaction \citep[see e.g.][]{2019BambicReynolds,2018Ehlert+,2017BourneSijacki,2017Weinberger+,2016YangReynoldsB,2015Reynolds+,2012Krause+,2004Omma+}. Due to the large dynamic range needed to capture the disparate scales associated with this problem, it is not possible to entirely self-consistently launch jets; some sub-grid prescription must be used instead. Many of these studies fix the jet power and direction, which does not allow for the formation of self-regulated feedback loop, making it hard to reconcile these results with the findings of GRMHD simulations.

AGN feedback in galaxy formation simulations has been widely considered \citep{2019Dave+, 2018Henden+, 2018Weinberger+, 2017McCarthy+, 2015Sijacki+, 2015Schaye+, 2015Crain+, 2012Dubois+, 2007Sijacki+}. The inclusion of supermassive black holes in these simulations has proved necessary in order to regulate the properties of massive galaxies, thus producing more realistic galaxy populations that are broadly consistent with observations. Due to resolution constraints, however, the jet energy must often be injected in the form of a thermal or kinetic dump, e.g. in off-centre bubbles or large-scale (bipolar) winds, which eliminates the requirement that the lobe inflation be followed, albeit at the cost of reduced physical accuracy. Zoom-in simulations of cosmological clusters somewhat address the resolution problems associated with full cosmological box simulations. Indeed, individual jet studies that follow the lobe inflation process in a live cosmological environment have been carried out \citep{2020BourneSijacki,2019Bourne+,2010Dubois+,2010Morsony+,2006Heinz+}. 

Despite these problems, the modelling of jet feedback in galaxy and cosmological scale simulations is becoming more sophisticated, incorporating more accurate modelling of (computationally) complex physical processes such as magnetic fields and cosmic rays \citep{2019Yang+, 2018Ehlert+, 2017Weinberger+}. Jet studies are also now finding that they are able to reproduce features seen in radio and X-ray observations with unprecedented accuracy \citep{2020BourneSijacki, 2019Bourne+}. To investigate these issues fully, however, self-regulated jet feedback models need to be considered \citep[see e.g.][]{2016YangReynoldsB, 2014LiBryan, 2011Gaspari+, 2007CattaneoTeyssier, 2006VernaleoReynolds}, but note that self-consistent models that link the evolution of gas angular momentum with the jet power are still in their infancy.

Due to the vast range of scales that are involved in tracking accretion flows and AGN jet propagation in galaxy-scale simulations, many of the launching processes (that can be captured to some extent in GRMHD simulations) fall below scales that are resolvable. Our work here focuses on galaxy-scale simulations but aims to (somewhat) bridge the gap between galaxy-scale and GRMHD simulations. Since we cannot resolve the jet launching scales we have developed a self-consistent sub-grid model for AGN feedback in the form of a Blandford-Znajek jet, motivated by the results of GRMHD simulations and general relativistic analytic predictions. Our model assumes that the black hole is surrounded by a sub-grid, thin $\alpha$-disc which modulates the accretion flow onto the black hole, allowing us to accurately follow both the evolution of the mass and angular momentum of the black hole. In such a model it is of vital importance to self-consistently evolve the system as a whole and hence the evolution of the black hole as a result of launching the jet is included in our model as well. 

Our model has been designed such that it can be employed in galaxy scale simulations with the ultimate aim of assessing whether self-consistent Blandford-Znajek jets can reproduce large-scale galaxy and cluster observables. This work, however, specifically focuses on resolving the parsec-scale interactions of jets with the cold dense circumnuclear gas found close to the black hole as well as the interactions of the jet with the surrounding hot CGM.

The structure of this paper is as follows. In Section~\ref{Sec: Theory} we motivate and develop the analytic equations that govern the evolution of our sub-grid model. Then in Section~\ref{Sec: Numerical} we describe how this is implemented into \textsc{arepo} and explain the refinement schemes we employ. In Section~\ref{Sec: Setup} we introduce the simulation setup that we use to test our sub-grid model. In Section~\ref{sec: fixed} we discuss simulations in which the jet power and direction are fixed which we then use to better understand the results of analogous simulations in which we utilise our full Blandford-Znajek jet model, which are presented in Section~\ref{sec: BZ results}. In Section~\ref{sec: discussion} we discuss our work in the context of previous galaxy-scale simulations and highlight any physics missing from our simulations and finally, we end with our conclusions in Section~\ref{sec: conclusions}.

\section{Theoretical background}
\label{Sec: Theory}
In our sub-grid prescription for black hole accretion and feedback, we model the black hole as a sink particle surrounded by a sub-grid accretion disc which follows the structure of a steady-state, geometrically-thin $\alpha$-disc, as presented in \cite{1973ShakuraSunyaev}. The black hole has a sub-resolution mass of $M_{\rm BH}$ and angular momentum $\mathbf{J}_{\rm BH}$. Similarly, the disc is characterised solely by its mass, $M_{\rm d}$, and total angular momentum, $\mathbf{J}_{\rm d}$. The sub-grid disc-black hole system interacts gravitationally with the wider simulation via its dynamical mass, $M_{\rm dyn} = M_{\rm BH} + M_{\rm d}$.

Our implementation of accretion via a thin, potentially warped $\alpha$-disc largely follows the procedure presented in \cite{2018Fiacconi+}. Our numerical implementation, however, is somewhat different (see Section~\ref{Sec: Numerical}) and the launching of the Blandford-Znajek jet necessarily changes the equations governing the evolution of the system. We now proceed to fully develop these evolutionary equations.

\subsection{Accretion and inflow}
\label{Subsec: Accretion and inflow}
We first determine how the black hole and disc will evolve due to inflow of material from the wider simulation onto the sub-grid disc and the flow of mass onto the black hole from the inner edge of the disc.

In the \cite{1973ShakuraSunyaev} model, the disc extends down the the innermost stable circular orbit (ISCO) of the black hole and accretion is driven by an effective viscosity, $\nu$, parameterised by $\alpha$, according to
\begin{equation}
    \nu \approx \alpha c_{\rm s} \, H\, , 
\end{equation}
where $c_{\rm s}$ is the sound speed and $H$ is the disc scale-height. Whilst the value of this $\alpha$-parameter is poorly constrained by theory, it enters into the structural equations raised only to low powers. We therefore choose to fix it at $0.1$, consistent with some observations \citep{2007King+, 2003Schreiber+, 2001Cannizzoa, 2001Cannizzob}.

The mass of the black hole evolves due to the accretion of material\footnote{Specifically the binding energy associated with the material.} from the inner edge of the $\alpha$-disc. If the rest-mass flux onto the black hole is $\dot{M}_{\rm{BH},0}$ then the corresponding increase in black hole mass will be
\begin{equation}
    \dot{M}_{\rm BH}^{\rm (acc)} = (1-\epsilon_{\rm r})\, \dot{M}_{\rm BH,0}\, ,
\end{equation}
where $\epsilon_{\rm r}$ is the spin dependent radiative efficiency which corrects for the reduced binding energy of material passing through the ISCO \citep{1973NovikovThorne} (see Appendix~\ref{App: Spin dep params} for the full expression for $\epsilon_{\rm r}$).

The mass of the accretion disc will evolve due to inflow of material from the surroundings as well as the loss of material at its inner edge. Assuming that the inflowing material circularises and settles into the disc, $M_{\rm d}$ will change according to
\begin{equation}
\label{eq: MdotDiscAcc}
    \dot{M}_{\rm d}^{\rm (acc)} = \dot{M}_{\rm in} -\dot{M}\, ,
\end{equation}
where $\dot{M}_{\rm in}$ is the mass inflow rate from the surroundings onto the black hole-disc system and $\dot{M}$ is the steady state rest mass flux through the disc.

In the above equations we have taken care to differentiate between the rest mass flux onto the black hole, $\dot{M}_{\rm BH,0}$ and the mass flux through the disc, $\dot{M}$. This allows for the possibility that not all material that leaves the disc at its inner edge is captured by the black hole. We do this as later we will assume that the jet is mass loaded with material from the accretion disc (see Section~\ref{Subsec: Jet Launching}).

The angular momentum direction of the accreted material is fixed by the assumption that the Bardeen-Petterson effect causes the inner disc to align with the equatorial plane of the black hole (as discussed in Section~\ref{Subsec: BP effect} below). This means that the angular momentum evolution of the black hole due to these mass flows is
\begin{equation}
    \dot{\mathbf{J}}_{\rm BH}^{\rm (acc)} = L_{\rm ISCO}\;\dot{M}_{\rm BH,0} \;\mathbf{j}_{\rm BH}\, ,
\end{equation}
where $L_{\rm ISCO}$ is the magnitude of the specific angular momentum of the gas that is accreted by the black hole from the disc (see Appendix~\ref{App: Spin dep params} for the full expression of $L_{\rm ISCO}$ in terms of the black hole spin) and $\mathbf{j}_{\rm BH}$ is the unit vector in the direction of the black hole angular momentum.

Similarly, the disc angular momentum evolution is given by
\begin{equation}
    \dot{\mathbf{J}}_{\rm d}^{\rm (acc)} = \dot{M}_{\rm in}\mathbf{L}_{\rm in} - L_{\rm ISCO}\dot{M} \,\mathbf{j}_{\rm BH}\; ,
\end{equation}
where $\mathbf{L}_{\rm in}$ is the specific angular momentum of the inflowing gas.

In accordance with the assumptions of the \cite{1973ShakuraSunyaev} thin disc equations we cap the mass flux through the disc, $\dot{M}$, at the Eddington rate, $\dot{M}_{\rm Edd}$. Parameterising $\dot{M}$ in terms a fraction $f_{\rm Edd}$ of $\dot{M}_{\rm Edd}$ we can write
\begin{align}
    \dot{M} &= {\rm min}(f_{\rm Edd}, 1) \; \dot{M}_{\rm Edd}\, ,\nonumber \\
    &= {\rm min}(f_{\rm Edd}, 1) \; \frac{M_{\rm BH}}{\tau_{\rm Salp}} \; ,
\end{align}
where $\tau_{\rm Salp}$ is the Salpeter time
\begin{equation}
    \tau_{\rm Salp} = \frac{\epsilon_{\rm r}\,\kappa_{\rm ES} c}{4\pi G} \approx 4.5 \times 10^7 \, \bigg(\frac{\epsilon_{\rm r}}{0.1}\bigg) \; {\rm yrs}  \; ,
\end{equation}
and $\kappa_{\rm ES}\approx 0.4 \; {\rm cm^2 g^{-1}}$ is the electron-scattering opacity. $f_{\rm Edd}$ is determined using the solutions to the \cite{1973ShakuraSunyaev} thin disc equations under the assumption that the disc is gas pressure dominated, in steady state and that electron-scattering dominates the opacity which altogether give
\begin{equation}
\label{eq: fedd}
    f_{\rm Edd} \approx 0.76 \bigg(\frac{\epsilon_{\rm r}}{0.1}\bigg)\bigg(\frac{M_{\rm d}}{10^4 \,{\rm M_\odot}}\bigg)^5\bigg(\frac{M_{\rm BH}}{10^6 \,{\rm M_\odot}}\bigg)^{-47/7}\bigg(\frac{a \; J_{\rm d} / J_{\rm BH}}{3}\bigg)^{-25/7}\, ,
\end{equation}
where $a$ is the dimensionless spin parameter
\begin{equation}
    a = \frac{c\,J_{\rm BH}}{G \, M_{\rm BH}^2}\, .
\end{equation}
Theory constrains $a$ to lie in the range $0\leq a\leq 1$, however, if the black hole is surrounded by a thin disc then preferential capture of negative angular momentum photons from the surface of the disc will cap the spin such that $a\leq0.998$ \citep{1974Thorne}.

At low accretion rates we expect radiative cooling to become inefficient and the disc to thicken \citep{1994NarayanYi, 1995NarayanYi, 1995Abramowicz+} and in this regime the \cite{1973ShakuraSunyaev} model breaks down. We therefore ensure that the accretion rate does not become too low by imposing $f_{\rm Edd} >f_{\rm Edd}^{\rm (min)} = 10^{-4}$, as in \cite{2018Fiacconi+}.

Equation~(\ref{eq: MdotDiscAcc}) admits the possibility that the accretion disc could grow without bound if $\dot{M}_{\rm in}$ is consistently higher than $\dot{M}$. Were this to be the case, the disc could enter a state in which self-gravity becomes important in its outer regions. This self-gravitating regime is highly non-linear and could lead to processes such as the formation of spiral arms and bars, fragmentation of the disc and, ultimately, star formation. This regime is poorly understood \citep{2003Goodman} and it is not known how these processes may couple to alter the mass and angular momentum transport through the disc. We therefore choose to prevent the disc from entering such a regime by capping the mass inflow rate such that the disc mass remains below the mass at which the material at the outer edge of the disc would be unstable according to the Toomre criterion
\begin{equation}
\label{eq: Msg}
    M_{\rm d}^{\rm(sg)} \approx 2.2\times 10^4 \; \bigg(\frac{f_{\rm Edd}}{\epsilon_{\rm r}/0.1}\bigg)^{4/45}\bigg(\frac{M_{\rm BH}}{10^6 \,{\rm M_\odot}}\bigg)^{34/45} \; {\rm M_\odot} \, .
\end{equation}
For the derivation of the above expression for $M_{\rm d}^{\rm (sg)}$, see appendix A of \cite{2018Fiacconi+}.

\subsection{The Bardeen-Petterson effect}
\label{Subsec: BP effect}
A spinning black hole leads to gas on non-equatorial orbits precessing about the angular momentum vector of the black hole at a rate that depends on the distance from the black hole, i.e. the Lense-Thirring effect \citep{1918LenseThirring}. This precession is countered by the vertical viscosity in the disc which, in the thin disc regime, can then lead to the disc becoming warped. The inner regions of an initially misaligned disc will (counter-)align with the spin of the black hole while the outer disc remains misaligned, the so-called Bardeen-Petterson effect \citep{1975BardeenPetterson}. The radius at which the transition occurs, the warp radius $R_{\rm warp}$, is determined by a balance of the vertical warp propagation timescale, $\tau_{\nu_2}$, and the Lense-Thirring precession timescale, $\tau_{\rm LT}$,
\begin{equation}
    \frac{R_{\rm warp}^2}{\nu_2} \equiv \tau_{\nu_2}(R_{\rm warp}) = \tau_{\rm LT}(R_{\rm warp}) \equiv \frac{c^2 R_{\rm warp}^3}{2 \, G \, J_{\rm BH}} \, ,
\end{equation}
where $\nu_2$ is the vertical viscosity \citep[see e.g.][]{2013Dotti+, 2007Martin+, 2006LodatoPringle}. 
In the steady, thin disc regime the warp will propagate outwards diffusively and, provided the warp radius lies within the disc, the disc can attain a steady, warped state since the warping timescale is shorter than the local viscous timescale. Gas in the misaligned outer disc will experience a torque from the black hole that is strongest in the region around the warp radius. In an isolated black hole-disc system, the total angular momentum, $\mathbf{J}_{\rm tot} = \mathbf{J}_{\rm BH} + \mathbf{J}_{\rm d} $ is conserved, meaning that an equal and opposite torque must be felt by the black hole itself in order that $\mathbf{J}_{\rm tot}$ is conserved. Altogether, this leads to precession and (counter-)alignment of the black hole and disc angular momentum vectors.

Since the torque exerted on the black hole due to the warping of the disc does not have a component in the direction of the black hole angular momentum \citep{1992Pringle} it can be expressed in the form 
\begin{equation}
\label{eq: jdot bh bp}
    \dot{\mathbf{J}}_{\rm BH}^{\rm (BP)} = -J_{\rm BH}\bigg\{K_1(\mathbf{j}_{\rm BH} \times \mathbf{j}_{\rm d}) + K_2\big[\mathbf{j}_{\rm BH} \times(\mathbf{j}_{\rm BH} \times \mathbf{j}_{\rm d})\big]\bigg\} \; ,
\end{equation}
where $\mathbf{j}_{\rm BH}$ and $\mathbf{j}_{\rm d}$ are the unit vectors in the direction of the black hole and disc angular momenta, respectively. The first term in equation~(\ref{eq: jdot bh bp}) corresponds to a precession about $\mathbf{J}_{\rm tot}$ and the second term leads to alignment with $\mathbf{J}_{\rm tot}$ \citep{2005King+}. In general $K_1$ and $K_2$ are arbitrary functions, however, under the assumptions that (i) the misalignment between the angular momenta is small and (ii)  the total angular momentum is dominated by that of the disc, we can use the results of \cite{2007Martin+} to identify $K_1$ and $K_2$ with the precession and alignment timescales. Whilst these assumptions are inherently restrictive, we find that this is the best option available to determine the Bardeen-Petterson torque, short of solving for the full structure of the accretion disc. We discuss the implications of this assumption in Section~\ref{sec: BZ results}. 

Ultimately, this allows us to express $K_1$ and $K_2$ as
\begin{equation}
\label{eq: K params}
    K_1 = \frac{\sin(\pi/7)}{\tau_{\rm GM}}  \qquad \text{and} \qquad K_2 = \frac{\cos(\pi/7)}{\tau_{\rm GM}}\, ,
\end{equation}
where $\tau_{\rm GM}$ is the gravitomagnetic timescale: 
\begin{equation}
    \tau_{\rm GM} \approx 0.17 \bigg(\frac{M_{\rm BH}}{10^6 \,{\rm M_\odot}}\bigg)^{-2/35}\bigg(\frac{f_{\rm Edd}}{\epsilon_{\rm r}/0.1}\bigg)^{-32/35} a^{5/7} \; \rm{Myrs} \; ,
\end{equation}
\citep{2013Dotti+, 2009Perego+, 2007Martin+}.

Equation~(\ref{eq: jdot bh bp}) and the identities in equation~(\ref{eq: K params}) together determine the evolution of the black hole angular momentum due to the Bardeen Petterson effect. The evolution of the disc is then simply
\begin{equation}
\label{eq: JdotDBP}
    \dot{\mathbf{J}}_{\rm d}^{\rm (BP)} =  - \dot{\mathbf{J}}_{\rm BH}^{\rm (BP)}\; .
\end{equation}

If the warp radius is larger than the disc extent, however, the disc will not reach a steady warped state and these equations for the angular momentum evolution are no longer valid. This will be the case if the mass of the black hole exceeds a characteristic warping mass\footnote{For the derivation of this warping mass see appendix A of \cite{2018Fiacconi+}.}
\begin{equation}
\label{eq: Mwarp}
    M_{\rm BH}^{\rm (warp)} \approx 10^7 \, \bigg(\frac{M_{\rm d}}{10^4\, {\rm M_\odot}}\bigg)^{35/82}\, \bigg( \frac{f_{\rm Edd}}{\epsilon_{\rm r}/0.1}\bigg)^{-17/82} \; a^{-25/82} \; {\rm M_\odot} \; .
\end{equation}
In this scenario the disc angular momentum will \mbox{(counter-)align} with that of the black hole on a timescale set by the diffusive warp propagation. In this regime we choose to employ a simple prescription and assume that the disc angular momentum instantaneously (counter-)aligns with that of the black hole. Alignment will occur if $\mathbf{j}_{\rm BH} \cdot \mathbf{j}_{\rm d} > -J_{\rm BH} / 2J_{\rm d}$ \citep{2005King+}, otherwise they will counter-align.

\subsection{The Blandford-Znajek mechanism}
\label{subsec: BZ}
\cite{1977BlandfordZnajek} proposed a mechanism through which magnetic fields could extract rotational energy from the black hole. A large poloidal magnetic field that threads a spinning black hole becomes twisted due to frame-dragging effects and a toroidal field develops. The resulting magnetic pressure in this configuration then provides the launching mechanism for the jet with the work done by the black hole on the field lines leading to the extraction of its rotational energy. 

\cite{1977BlandfordZnajek} determined the perturbations to the fields under the assumption that the black hole is surrounded by a razor-thin disc and its magnetosphere is force-free. If the magnetic flux threading one hemisphere of the black hole horizon is
\begin{equation}
    \Phi_{\rm BH} = \frac{1}{2}\int \abs{B^{r}} \D A_{\theta\phi}\, ,
\end{equation}
where $B^{r}$ is the radial component of the magnetic field and $\D A_{\theta\phi}$ is the area element on the horizon then, in the limit of $a\ll 1$ and with fixed $\Phi_{\rm BH}$, the rotational energy of the black hole can be extracted at a rate
\begin{equation}
\label{eq: Energy flux BZ}
    \dot{E}_{\rm BZ} = \frac{\kappa}{4\pi}\Phi_{\rm BH}^2\frac{a^2\, c}{16 r_{\rm g}^2}\, ,
\end{equation}
where $r_{\rm g}=G\,M_{\rm BH}/c^2$ is the black hole gravitational radius and $\kappa$ is a parameter that depends only weakly on the magnetic field geometry. In this work we fix $\kappa = 1/(6\pi)$ which corresponds to a split-monopole geometry.

Launching a Blandford-Znajek jet leads to a decrease in the mass-energy of the black hole and causes it to spin-down. The energy flux due to the Blandford-Znajek effect has been expanded analytically to higher orders under various assumptions \citep{2010Tchekhovskoy+, 2008TanabeNagataki} and there have been many studies investigating the process via GRMHD simulations \citep{2013Penna+, 2012McKinney+, 2006HawleyKrolik, 2004McKinneyGammie, 2001Komissarov}. In Appendix~\ref{App: Derivation} we provide an explicit derivation of the energy and angular momentum fluxes to second and first order in the perturbed fields, respectively. When implementing our sub-grid model, however, we also include the higher order terms in the energy flux determined numerically by \cite{2010Tchekhovskoy+} as this improves the accuracy of the determination of the energy flux at higher spin values. We can therefore write the energy flux as 
\begin{equation}
    \dot{E}_{\rm BZ} = \frac{\kappa}{4\pi}\,\phi_{\rm BH}^2\,  f(a) \, \dot{M}_{\rm BH,0} \, c^2\, ,
\end{equation}
where the dimensionless magnetic flux is
\begin{equation}
    \phi_{\rm BH} = \frac{\Phi_{\rm BH}}{\sqrt{\dot{M}_{\rm BH, 0} \, r_{\rm g}^2 c}} \, ,
\end{equation} 
and $f(a)$ is a dimensionless function of the black hole spin, given by
\begin{align}
    f(a) = \Bigg(\frac{a}{2\,\big(1+ \sqrt{1-a^2}\big)}\Bigg)^2 + 1.38\,&\Bigg(\frac{a}{2\,\big(1+ \sqrt{1-a^2}\big)}\Bigg)^4\nonumber \\
    -9.2\,&\Bigg(\frac{a}{2\,\big(1+ \sqrt{1-a^2}\big)}\Bigg)^6 \, .
\end{align}

This outward flux of energy leads to the evolution of black hole mass according to 
\begin{equation}
   \dot{M}_{\rm BH}^{\rm (BZ)} =  - \epsilon_{\rm BZ} \dot{M}_{\rm BH,0} c^2 \, ,
\end{equation}
where, in analogy with the radiative efficiency, we have introduced a `Blandford-Znajek efficiency'
\begin{equation}
    \epsilon_{\rm BZ} \equiv \frac{\kappa}{4\pi}\phi_{\rm BH}^2 f(a) \, .
\end{equation}

The corresponding outward angular momentum flux leads to evolution of the black hole angular momentum
\begin{equation}
    \dot{\mathbf{J}}_{\rm BH}^{\rm (BZ)} = -\dot{M}_{\rm BH, 0} \, \mathbf{L}_{\rm BZ} \, ,
\end{equation}
which we have defined in terms of an effective specific angular momentum
\begin{align}
    \mathbf{L}_{\rm BZ} &= L_{\rm BZ}\,\mathbf{j}_{\rm BH}\; , \nonumber \\
    &=\frac{\kappa}{2\pi}\,\phi_{\rm BH}^2\,\Bigg(\frac{a}{2\,\big(1+ \sqrt{1-a^2}\big)}\Bigg)\,\frac{G\,M_{\rm BH}}{c}\,\mathbf{j}_{\rm BH} \, .
\end{align}
The dimensionless magnetic flux that threads the black hole horizon, $\phi_{\rm BH}$, cannot be directly measured from our simulations since we do not track magnetic fields (and even if we did, the processes that determine it would be well below our resolution scale). We therefore choose to use the spin dependent values for $\phi_{\rm BH}$ obtained from the GRMHD simulations in \cite{2012Tchekhovskoy}. It should be noted here that these simulations correspond to an accretion disc in a magnetically arrested state (i.e. the magnetic flux on the black hole has reached saturation). We will discuss this choice further in Section~\ref{sec: discussion}.

\subsection{Jet launching}
\label{Subsec: Jet Launching}
When launching the Blandford-Znajek jet we explicitly mass load it by assuming that some of the mass that flows through the ISCO of the sub-grid accretion disc is drawn up into the jet. We control the fraction of this mass flux that goes into the jet using the mass-loading factor, $\eta_{\rm J}$ \citep{2017BourneSijacki, 2010Ostriker+}, a free parameter, which we define to be
\begin{equation}
    \dot{M}_{\rm J} = \frac{\eta_{\rm J}}{1+\eta_{\rm J}}\dot{M} \, .
\end{equation}
Throughout this work we choose $\eta_{\rm J} = 1$ \citep{2017BourneSijacki}. 

In terms of the mass-loading factor, we can express the rest mass flux across the black hole horizon as
\begin{equation}
    \label{Eq: Mdot BH rest}
    \dot{M}_{\rm BH, 0} = \frac{1}{1+\eta_{\rm J}}\dot{M}\, .
\end{equation}
The energy and momentum fluxes into the jet are
\begin{align}
    \label{eq: energy jet}
   \dot{E}_{\rm J} &= \frac{1}{2}\dot{M}_{\rm J} v_{\rm J}^2 \, ,\\
   \label{eq: momentum jet}
   \dot{P}_{\rm J} &= \dot{M}_{\rm J} v_{\rm J}\, ,
\end{align}
where $v_{\rm J}$ is the sub-resolution velocity of the jet.

We assume that the jet is powered by the entire outward flux of energy from the black hole, as predicted by the Blandford-Znajek process, and thus
\begin{align}
    \label{eq: edot flux}
   \dot{E}_{\rm J} &= \epsilon_{\rm BZ} \dot{M}_{\rm BH, 0} c^2 \, ,\\
   \label{eq: pdot flux}
   \dot{P}_{\rm J} &=  \sqrt{2\dot{E}_{\rm J}\dot{M}_{\rm J}} = \frac{\sqrt{2\eta_{\rm J}\, \epsilon_{\rm BZ}}}{1+\eta_{\rm J}}\dot{M} c \, .
\end{align}

It should be noted here that in our numerical scheme we inject the jet energy into a finite mass. Since this leads to numerical mass-loading, the energy and momentum fluxes into the jet cannot satisfy equations~(\ref{eq: edot flux})~and~(\ref{eq: pdot flux}) simultaneously. We choose to conserve kinetic energy and so the momentum flux in equation~(\ref{eq: pdot flux}) will not correspond to the momentum flux into the jet and similarly the sub-resolution velocity, $v_{\rm J}$, will not correspond to the initial velocity of jet material in the simulations. The specifics of the jet injection are detailed in Section~\ref{subsec: Jet injection} below.

\subsection{Summary of equations}
\label{Sec: Summary}
We end this section by collating the relevant equations, developed above, into a set of overarching equations that govern the evolution of the black hole and accretion disc and the properties of the jet in our model.

The black hole mass evolves according to
\begin{align}
    \dot{M}_{\rm BH} &= \dot{M}_{\rm BH}^{\rm (acc)}+ \dot{M}_{\rm BH}^{\rm (BZ)} \, ,\\
    \label{eq: MdotBH}
    &=\frac{(1-\epsilon_{\rm r}-\epsilon_{\rm BZ})}{1 + \eta_{\rm J}}\dot{M} \, ,
\end{align}
and the disc mass
\begin{align}
\label{eq: MdotDisc}
    \dot{M}_{\rm d} = \dot{M}_{\rm in} -\dot{M} \, .
\end{align}
The black hole angular momentum is given by 
\begin{align}
    \dot{\mathbf{J}}_{\rm BH} =& \,\dot{\mathbf{J}}_{\rm BH}^{\rm (acc)} + \dot{\mathbf{J}}_{\rm BH}^{\rm (BZ)} + \dot{\mathbf{J}}_{\rm BH}^{\rm (BP)} \, ,\\
    \label{eq: jdotbh}
    =&\,(L_{\rm ISCO}- L_{\rm BZ})\,\frac{1}{1+\eta_{\rm J}}\dot{M}\, \mathbf{j}_{\rm BH}\nonumber \\ &-J_{\rm BH}\bigg\{\frac{\sin(\pi/7)}{\tau_{\rm GM}}( \mathbf{j}_{\rm BH} \times \mathbf{j}_{\rm d}) + \frac{\cos(\pi/7)}{\tau_{\rm GM}} \big[\mathbf{j}_{\rm BH} \times(\mathbf{j}_{\rm BH} \times \mathbf{j}_{\rm d})\big]\bigg\} \, ,
\end{align}
and that of the disc is given by
\begin{align}
    \dot{\mathbf{J}}_{\rm d} =& \,\dot{\mathbf{J}}_{\rm d}^{\rm (acc)} - \dot{\mathbf{J}}_{\rm BH}^{\rm (BP)} \, ,\\
    \label{eq: jdotd}
    =& \,\dot{M}_{\rm in}\,\mathbf{L}_{\rm in} - L_{\rm ISCO}\,\dot{M} \,\mathbf{j}_{\rm BH}\nonumber\\ &+J_{\rm BH}\bigg\{\frac{\sin(\pi/7)}{\tau_{\rm GM}}( \mathbf{j}_{\rm BH} \times \mathbf{j}_{\rm d}) + \frac{\cos(\pi/7)}{\tau_{\rm GM}} \big[\mathbf{j}_{\rm BH} \times(\mathbf{j}_{\rm BH} \times \mathbf{j}_{\rm d})\big]\bigg\} \, .  
\end{align}

In terms of $\dot{M}$, the energy, mass and momentum fluxes into the jet are
\begin{equation}
\label{eq: EdotJet}
    \dot{E}_{\rm J} = \frac{\epsilon_{\rm BZ}}{1 + \eta_{\rm J}}\dot{M} c^2 \, ,
\end{equation}
\begin{equation}
\label{eq: PdotJet}
    \dot{P}_{\rm J} = \frac{\sqrt{2\eta_{\rm J}\epsilon_{\rm BZ}}}{1+\eta_{\rm J}}\,\dot{M}\, c \, ,
\end{equation}
\begin{equation}
\label{eq: MdotJet}
    \dot{M}_{\rm J} = \frac{\eta_{\rm J}}{1+\eta_{\rm J}}\dot{M} \, .
\end{equation}

\section{Numerical implementation}
\label{Sec: Numerical}
\begin{figure*}
    \centering
    \includegraphics[width=\textwidth]{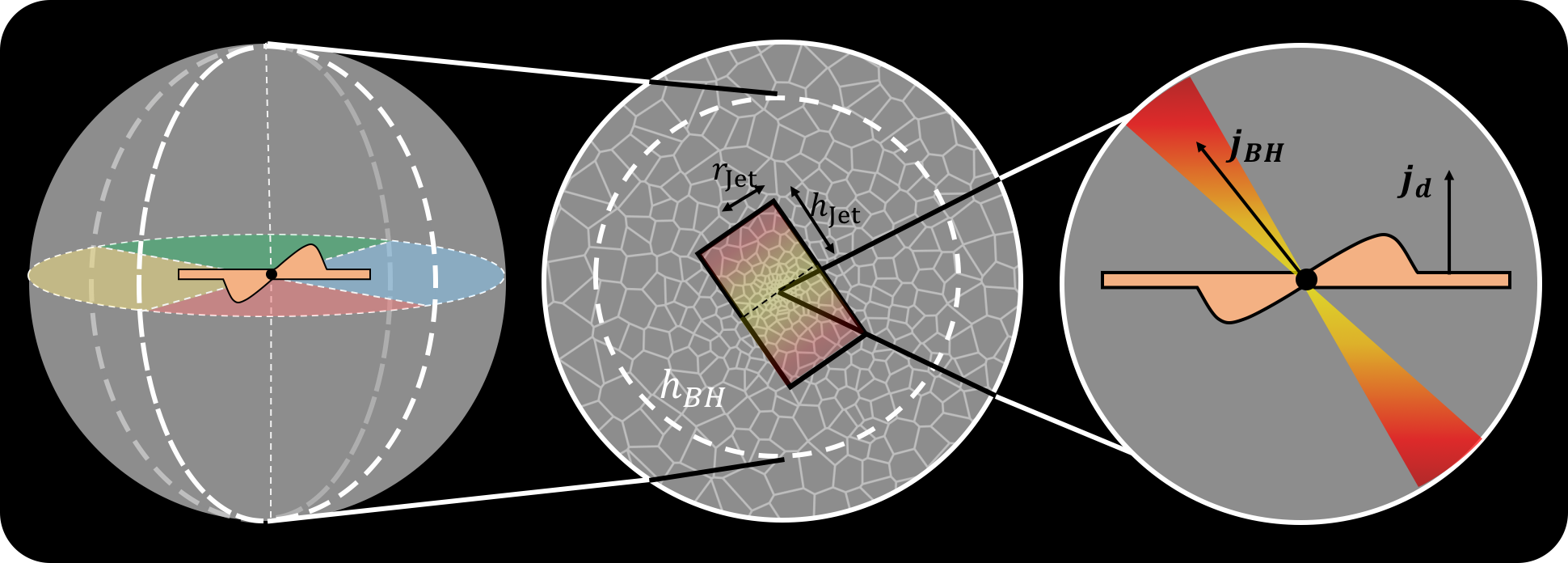}
    \caption{A schematic overview of our sub-grid Blandford-Znajek jet model. On the left we show how the sphere defined by the black hole smoothing length is split into four sectors based on the direction of the $\alpha$-disc angular momentum. The middle circle illustrates the jet injection in the context of the wider simulation. The cylinder into which the jet is injected is outlined in black and the white dashed line indicates the black hole smoothing length. The circle on the right shows the angular momenta of the black hole and warped $\alpha$-disc and how these correspond to the jet direction.}
    \label{fig: schematic}
\end{figure*}

In the previous section we described the physical processes that govern our sub-grid model and determined the equations that describe the evolution of the system. In this section we detail how we have implemented this theoretical model in our numerical framework.
\subsection{AREPO}
We have incorporated this new black hole accretion and feedback prescription into the moving-mesh code \textsc{arepo} \citep{2010Springel, 2016Pakmor+, 2020Weinberger+}. 
\textsc{arepo} solves the equations of hydrodynamics on a quasi-Lagrangian, moving, unstructured mesh based on the Voronoi tessellation of a set of points which move with the fluid with small corrections to ensure regularisation of the cells. The code therefore inherits many benefits of Lagrangian methods such as Galilean invariance and density dependent resolution whilst also maintaining the high accuracy treatment of shocks and instabilities that is characteristic of Eulerian codes. The gravitational forces are calculated using a hierarchical oct-tree method \citep{2005Springel, 1986BarnesHut} along with the PM method for long-range forces.

\subsection{Gas inflow properties}
\label{Sec: Gas inflow props}
At the beginning of each black hole timestep we first estimate the gas inflow rate, $\dot{M}_{\rm in}$, and the specific angular momentum of this gas, $\mathbf{L}_{\rm in}$, using a smooth particle hydrodynamic interpolation of the local mass flux onto black hole, calculated within the black hole smoothing length, $h_{\rm BH}$\footnote{For discussion of the convergence of the inflow rate estimate with respect to $h_{\rm BH}$ we direct the reader to appendix C of \cite{2018Fiacconi+}.}.

We do so by splitting the sphere defined by this smoothing length into four sectors along the direction parallel to the angular momentum vector of the $\alpha$-disc (as shown in Fig.~\ref{fig: schematic}). The reasoning behind splitting the sphere into sectors will be explained shortly. The mass inflow rate is then calculated for each sector individually according to
\begin{equation}
    \dot{M}_{{\rm in}, s} = -\int_{\mathcal{A}_s} \rho \mathbf{u}\cdot \D\mathbf{S} \approx -\mathcal{A}_s\mathcal{F}_{{\rm BH}, s} \, ,
\end{equation}
where $s\in \{1,2,3,4\}$ indexes the sector, $\rho$ and $\mathbf{u}$ are the density and velocity fields, respectively, and $\D\mathbf{S}$ is the outward area element. $\mathcal{F}_{{\rm BH}, s}$ is the numerical estimate of the mass flux evaluated at the position of the black hole for this sector and $\mathcal{A}_s$ is an effective area. 

We approximate the mass flux in each sector by summing over all $N_s$ gas cells that are radially inflowing and located within the relevant sector
\begin{equation}
    \mathcal{F}_{{\rm BH}, s} = \frac{\sum_{j=1}^{N_s}\rho_j u_{r,j}\, W(d_j)}{\sum_{j=1}^{N_s}\, W({d_j)}} \, ,
\end{equation}
where $\rho_j$ is the mass density of the cell, $u_{r,j} = (\mathbf{r}_j - \mathbf{r}_{\rm BH})\cdot(\mathbf{u}_j - \mathbf{u}_{\rm BH})/\abs{\mathbf{r}_j - \mathbf{r}_{\rm BH}}$ is the radial velocity of the cell and $W(d_j)$ is a cubic spline kernel with compact support over $h_{\rm BH}$ and $d_j = \abs{\mathbf{r}_j - \mathbf{r}_{\rm BH}}/h_{\rm BH}$.

We define the effective area to be $\mathcal{A}_s \equiv \pi r^2_{{\rm in},s}$ where $r_{{\rm in},s}$ is a characteristic inflow radius calculated for each sector
\begin{equation}
   r_{{\rm in}, s} = \frac{\sum_{j=1}^{N_s}(3V_j/4\pi)^{1/3}\,W(d_j)}{\sum_{j=1}^{N_s}\,W(d_j)}\, ,
\end{equation}
where $V_j$ is the volume of the gas cell.

Similarly, the specific angular momentum of the inflowing gas in each sector is given by
\begin{equation}
\label{eq: lin s}
    \mathbf{L}_{{\rm in}, s}\equiv \frac{\dot{\mathbf{J}}_{{\rm in}, s}}{\dot{M}_{{\rm in}, s}} = \frac{\sum_{j=1}^{N_s}\rho_j u_{r,j}\,\mathbf{L}_j\,W(d_j)}{\sum_{j=1}^{N_s}\rho_j u_{r,j}\,W({d_j)}} \, ,
\end{equation}
where $\mathbf{L}_j$ is the specific angular momentum of the cell.

We expect that inflowing gas will only be able to circularise and settle in the disc if it has specific angular momentum that is smaller than that of the outer edge of the disc
\begin{equation}
    L_{{\rm in}, s} < L_{\rm d}(R_{\rm out}) \approx \frac{J_{\rm d}}{M_{\rm d}} \, .
\end{equation}
We therefore check each sector (that contains inflowing material) individually to ensure this criterion is met. If this condition is not satisfied then we assume that there is no inflow in this sector and set $\dot{M}_{{\rm in}, s} = 0$. 

Carrying out this check for each sector individually allows us to better model non-axisymmetric feeding of the black hole via coherent streams of material. If accretion were to proceed in such a way, then calculating $\mathbf{L}_{\rm in}$ using equation~(\ref{eq: lin s}), but with the sum over all cells within $h_{\rm BH}$ may lead to the circularisation condition failing in scenarios where we would expect the these streams to individually be able to reach the disc, due to the nature of vector addition. We chose to use four sectors as we found that this gave significant improvement in the accuracy of our inflow estimates whilst still ensuring that each sector is sufficiently populated and that the associated numerical overhead is minimised\footnote{We have verified that using $2$-$16$ sectors in this calculation gives consistent estimates for the inflow properties.}.

After this circularisation check has been carried out we are then able to combine the values of $\dot{M}_{{\rm in}, s}$ and $\mathbf{L}_{{\rm in}, s}$ to get a global values for the mass inflow rate, $\dot{M}_{\rm in}$, and the specific angular momentum of inflowing material, $\mathbf{L}_{\rm in}$, by summing over all sectors, with appropriate weightings. We store the values of $\dot{M}_{{\rm in}, s}$ for each sector to ensure that we drain the appropriate amount of gas from each sector later in the timestep.

We now perform a series of further checks on $\dot{M}_{\rm in}$ and $\mathbf{L}_{\rm in}$ to ensure that the assumptions of our model are satisfied. We first verify that the specific angular momentum of the outer edge of the $\alpha$-disc is larger than that required to maintain a circular orbit at the ISCO (which holds true at all times in all simulations presented in this work).

At this point we also cap $\dot{M}_{\rm in}$ to ensure that the predicted mass of the accretion disc at the end of this timestep does not exceed the mass at which we expect it to become self-gravitating (see equation~(\ref{eq: Msg})). Finally, we limit $\dot{M}_{\rm in}$ to ensure that the disc remains sub-Eddington by linearly interpolating equation~(\ref{eq: fedd}) to ensure $f_{\rm Edd} < 1$. If any of these checks lead to $\dot{M}_{\rm in}$ being reduced then we also decrease the mass inflow rate in each sector (where $\dot{M}_{{\rm in}, s}\neq 0$) by the relevant amount. 

\subsection{Black hole and disc mass evolution}
\label{Sec: BH and disc mass evolution}

The black hole and disc masses are then evolved by integrating equations~(\ref{eq: MdotBH})~and~(\ref{eq: MdotDisc}) over the timestep using a second-order predictor-corrector Heun's scheme. Letting the subscript `n' indicate a value at the beginning of the time-step; `temp', a value after the predictor step and `n+1', a value after the corrector step, this integration procedure corresponds to
\begin{align}
    &M_{\rm BH,temp} = M_{\rm BH,n} + \bigg(\frac{1-\epsilon_{\rm r}-\epsilon_{\rm BZ}}{1 + \eta_{\rm J}}\bigg) \frac{f_{\rm Edd}}{\tau_{\rm Salp}}\, M_{\rm BH,n}\Delta t \; , \nonumber \\
    &M_{\rm d, temp} = M_{\rm d, n} - \frac{f_{\rm Edd}}{\tau_{\rm Salp}}\,M_{\rm BH,n}\Delta t + \dot{M}_{\rm in} \Delta t \; ,\nonumber \\
    \vspace{5pt}
    &M_{\rm BH,n+1} = M_{\rm BH, n} + \bigg(\frac{1-\epsilon_{\rm r}-\epsilon_{\rm BZ}}{1 + \eta_{\rm J}}\bigg) \frac{f_{\rm Edd}}{\tau_{\rm Salp}} \bigg(\frac{M_{\rm BH,n} + M_{\rm BH,temp}}{2}\bigg) \Delta t \; ,\nonumber \\
    &M_{\rm d, n+1} = M_{\rm d, n} - \frac{f_{\rm Edd}}{\tau_{\rm Salp}} \bigg(\frac{M_{\rm BH,n} + M_{\rm BH,temp}}{2}\bigg)\Delta t + \dot{M}_{\rm in} \Delta t \; .
\end{align}

\subsection{Black hole and disc angular momentum evolution} 
The angular momenta of the black hole and disc evolve according to equations~(\ref{eq: jdotbh})~and~(\ref{eq: jdotd}). We implement this numerically by splitting the calculation into two steps. In the first step we evolve the $\alpha$-disc due to inflow from the surroundings and the loss of material that mass-loads the jet from the ISCO. Simultaneously we evolve the black hole angular momentum according to predictions of the Blandford-Znajek jet model
\begin{equation}
\label{eq: jdot d temp}
    \mathbf{J}_{\rm d, temp} = \mathbf{J}_{\rm d, n} + \mathbf{L}_{\rm in} \dot{M}_{\rm in}\Delta t - \frac{\eta_{\rm J}}{1+\eta_{\rm J}} L_{\rm ISCO}\dot{M}\Delta t \;\mathbf{j}_{\rm BH,n} \; ,
\end{equation}
\begin{equation}
\label{eq: jdot bh temp}
    \mathbf{J}_{\rm BH, temp} = \mathbf{J}_{\rm BH, n} - \frac{1}{1+\eta_{\rm J}} L_{\rm BZ}\;\dot{M} \;\Delta t\;\mathbf{j}_{\rm BH,n} \, ,
\end{equation}
where we are now using the subscript `temp' to indicate the angular momenta after this step. 
The factor of $\eta_{\rm J}/(1+\eta_{\rm J})$ in equation~(\ref{eq: jdot d temp}) comes about due to the fact that we are only considering the material that is drawn up into the jet from the inner edge of the disc, $\eta_{\rm J} / (1+\eta_{\rm J})\;\dot{M}\;\Delta t$. Evolution due to the mass that flows onto the black hole from the inner edge of the disc, $1/(1+\eta_{\rm J})\;\dot{M}\;\Delta t$, is included in the subsequent step. We choose to break down the calculations in this way because the total angular momentum of the system after equations~(\ref{eq: jdot d temp})~and~(\ref{eq: jdot bh temp}) have been evaluated, $\mathbf{J}_{\rm cons} = \mathbf{J}_{\rm BH, temp}+\mathbf{J}_{\rm d, temp}$ is now conserved.

Having determined $\mathbf{J}_{\rm cons}$, we now begin the second step of the calculation and evolve the disc and black hole due to the accretion of material from the inner edge of the disc and due to mutual Bardeen-Petterson torques.

The Bardeen-Petterson effect does not alter the magnitude of the black hole angular momentum and also predicts that the inner disc will be (counter-)aligned with the spin of the black hole. This means that the final magnitude of the black hole angular momentum is affected by accretion alone
\begin{equation}
    J_{\rm BH, n+1} = J_{\rm BH, temp} + \frac{1}{1 + \eta_{\rm J}} L_{\rm ISCO} \dot{M} \Delta t \, .
\end{equation}
From this we calculate the predicted final spin of the black hole and cap it to 0.998 if necessary.

Then we check to see if the black hole mass exceeds the warping mass (see equation~(\ref{eq: Mwarp})). If not then the black hole angular momentum direction is evolved using a predictor-corrector method according to 
\begin{align}
    \dot{\mathbf{j}}_{\rm BH, n+1} = &-K_1(\mathbf{j}_{\rm BH, temp} \times \mathbf{j}_{\rm d, temp})\nonumber\\
    &-K_2\big[\mathbf{j}_{\rm BH, temp} \times(\mathbf{j}_{\rm BH, temp} \times \mathbf{j}_{\rm d, temp})\big] \; ,
\end{align}
which applies the Bardeen-Petterson torque. If, however, the black hole mass does exceed the warping mass then we assume that the angular momentum directions of the black hole and disc align instantaneously with that of the total angular momentum, as discussed in Section~\ref{Subsec: BP effect}.

Finally, the disc angular momentum is updated to satisfy total angular momentum conservation
\begin{equation}
    \mathbf{J}_{\rm d, n+1} = \mathbf{J}_{\rm cons} - \mathbf{J}_{\rm BH,n+1} \, .
\end{equation}
This final step incorporates the evolution of the disc due to the Bardeen-Petterson effect and due to the disc material accreted by the black hole.

\subsection{Mass draining}
\subsubsection{Sector based draining}
Mass is then drained from each gas cell in the simulation that has been determined to have contributed to the inflow onto the black hole-disc system during this timestep, as described in Section~\ref{Sec: Gas inflow props}. For a cell in a sector that has inflow and is itself radially inflowing, we drain mass
\begin{equation}
    \Delta m_{i} = {\rm max} \Bigg(\frac{m_{i}}{\sum_{j=0}^{N_s} m_{j}} \; \Delta M_{{\rm in},s}\;, \;0.99\,m_i\Bigg) \, ,
\end{equation}
where $i$ indexes the properties associated with an individual gas cell and $\Delta M_{{\rm in},s} = \dot{M}_{{\rm in},s} \, \Delta t$.

\subsubsection{Black hole dynamical mass evolution}
In the simulation, the dynamical mass associated with the black hole particle corresponds to that of the black hole-disc system as a whole, i.e the sum of the black hole mass and the accretion disc mass. We update the dynamical mass in accordance with this.

It should be highlighted here that black hole mass evolution is driven by fluxes of {\it binding energy} associated with gas that falls across its horizon and {\it not} the rest mass of this gas. This means that, in accordance with the predictions of general relativity, black hole accretion should not explicitly conserve mass, but rather energy. This lack of rest mass conservation is particularly obvious for our sub-grid model in situations where there is no net inflow of gas onto the system. In this scenario, no gas would be drained from the simulation but the dynamical mass of the system may change due to the accretion of material onto the black hole from the $\alpha$-disc.

\subsection{Jet injection}
\label{subsec: Jet injection}
We implement jet injection using a procedure similar to that of the `kinetic energy conserving jet' presented in \cite{2017BourneSijacki} which has been adapted for our Blandford-Znajek model.

We inject the jet, on scales that can be resolved, into a cylinder of radius $r_{\rm Jet}$ and height $2\,h_{\rm Jet}$ that is centred on the black hole. The cylinder has axis that lies along the direction of the jet (i.e. the direction of the black hole spin). $r_{\rm Jet}$ varies such that the total mass within the jet cylinder, $M_{\rm cyl}$, is constant\footnote{We ensure a minimum number of gas cells in each half of the jet cylinder meaning that the cylinder mass could be larger than this target mass. In Section~\ref{subsubsec: jet ref} we outline the refinement techniques employed to ensure the jet cylinder is sufficiently populated.}. Throughout this work we choose our target jet cylinder mass to be $M_{\rm cyl} = 0.5\;{\rm M_\odot}$. $h_{\rm Jet}$ is determined by $r_{\rm Jet}$ such that the aspect ratio of the cylinder is fixed, giving a constant jet opening angle, $\theta_{\rm Jet}$, that satisfies
\begin{equation}
    \tan \bigg(\frac{\theta_{\rm Jet}}{2}\bigg) =\frac{r_{\rm Jet}}{h_{\rm Jet}} \, .
\end{equation}
Following on from \cite{2017BourneSijacki}, we choose $r_{\rm Jet}/h_{\rm Jet} = 3/2$ which means that the cylinder has total volume $(4\pi/3)\ r_{\rm Jet}^3$. The cylinder is split into two halves, the north and the south (each with height $h_{\rm Jet}$), and half of the required integrated energy, mass and momentum flux is injected into gas cells in the `north' and the other half into those cells in the `south' (see Fig.~\ref{fig: schematic}). The increase in mass of a cell in the jet cylinder is given by
\begin{equation}
    \Delta m_i = \frac{1}{2}\dot{M}_{\rm J}\Delta t \; \frac{m_i\; W_{\rm Jet}(r_i, z_i)}{M_{\rm weight}} \, ,
\end{equation}
where $m_i$ is the initial mass of the cell, $\dot{M}_{\rm J}$ is as determined by equation~(\ref{eq: MdotJet}), the factor of a half accounts for the two halves of the jet cylinder and $M_{\rm weight}$ is the weighted sum of the masses of all gas cells in the relevant half-cylinder of the jet
\begin{equation}
    M_{\rm weight} = \sum^{\rm half-cyl}_{j} m_j\; W_{\rm Jet}(r_j, z_j) \, ,
\end{equation}
where $r_i$ is the (cylindrical) distance of the gas cell from the jet axis and $z_i$ is its height above the black hole equatorial plane. The kernel function has the form
\begin{equation}
    W_{\rm Jet}(r, z) \propto \exp{\Bigg(-\frac{r^2}{2r_{\rm Jet}^2}\Bigg)}\abs{z} \, .
\end{equation}

Momentum injection is carried out by giving gas cells a kick along the jet axis with magnitude
\begin{equation}
\label{eq: delta mom jet}
   \Delta p_i = \sqrt{2(m_{i} + \Delta m_i)(E^{\rm kin}_{i} + \Delta E_i^{\rm tot})} - \abs{\mathbf{p}_i} \, ,
\end{equation}
where $\mathbf{p}_i$ and $E^{\rm kin}_i$ are the initial momentum and kinetic energy of the cell and $\Delta E^{\rm tot}_i$ is the desired change in energy of the cell
\begin{equation}
    \Delta E^{\rm tot}_i = \frac{1}{2} \dot{E}_{\rm J} \,\Delta t \; \frac{m_i\; W_{\rm Jet}(r_i, z_i)}{M_{\rm weight}} \, ,
\end{equation}
where $\dot{E}_{\rm J}$ is as in equation~(\ref{eq: EdotJet}).

This single momentum kick, however, does not necessarily ensure that the total energy of the cell changes by $\Delta E_i^{\rm tot}$, as required by energy conservation, unless the momentum vector of the cell was initially is parallel to the jet axis. We therefore inject thermal energy, $\Delta \, E_i^{\rm therm}$, to ensure energy conservation, where
\begin{equation}
    \Delta \, E_i^{\rm therm} = E^{\rm kin}_i + \Delta E_i^{\rm tot} -  \frac{1}{2}\frac{\abs{\mathbf{p}_i + \Delta p_i\,\mathbf{j}_{\rm BH}}^2}{(m_i + \Delta m_i)} \, .
\end{equation}

It should be noted here that the Blandford-Znajek model predicts that the black hole spins down as the jet is launched. In order to conserve angular momentum we should also be giving the cells in the jet a kick in the toroidal direction. We neglect this effect, since we expect it to correspond to momentum kicks that are significantly smaller than those predicted by equation~(\ref{eq: delta mom jet}).

\subsection{Refinement}
\label{Subsec: Refinement}
Our model makes considerable use of \textsc{arepo}'s ability to refine and de-refine cells based on nearly arbitrary criteria which allows for high resolution in specific regions of interest whilst maintaining lower resolution where not needed. We employ refinement schemes that specifically target the region close to the black hole, the jet injection cylinder and the jet lobes. Together these allow us to accurately track the mass and angular momentum flows in the vicinity of the black hole and follow the propagation of the jet at sufficiently high resolution.

\subsubsection{Black hole refinement}
To obtain an accurate estimation the mass and angular momentum inflow rates, we make significant use of the super-Lagrangian refinement scheme outlined in \cite{2015CurtisSijacki} which adaptively increases the mass and spatial resolution in the region surrounding the black hole.

Within a chosen refinement radius, $R_{\rm ref}$, cells are forced to have radii\footnote{When we refer to the cell radius, we mean the radius of a sphere with volume equal to the cell volume: $R_i = \big(3V_i / 4\pi\big)^{1/3}$.}, $R_{i}$, between two values, $R_{\rm min}(d)$ and $R_{\rm max}(d)$ which increase linearly with distance from the black hole, $d$. Specifically, for a cell with $d_i \leq R_{\rm ref}$, we ensure 
\begin{equation}
    R_{\rm min}(d_i) < R_i < R_{\rm max}(d_i) \; ,
\end{equation}
where
\begin{align}
    R_{\rm min}(d) = \frac{d}{R_{\rm ref}}\frac{\big(R_{\rm cell}^{\rm max} - R_{\rm cell}^{\rm min}\big)}{C} + \frac{R_{\rm cell}^{\rm min}}{C} \; , \\
    R_{\rm max}(d) = \frac{d}{R_{\rm ref}}\big(R_{\rm cell}^{\rm max} - R_{\rm cell}^{\rm min}\big) + R_{\rm cell}^{\rm min} \; .
\end{align}
Here, $C$, $R_{\rm cell}^{\rm max}$, and $R_{\rm cell}^{\rm min}$ are free parameters to be specified. It should be noted that our choices for these parameters differ from those suggested in \cite{2015CurtisSijacki} as theirs were chosen with the aim of resolving the Bondi radius. The simulations we perform here are at a significantly higher spatial resolution and we are often resolving the Bondi radius before this super-Lagrangian refinement is applied.

\subsubsection{Jet refinement}
\label{subsubsec: jet ref}
Following the propagation of a jet at sufficiently high resolution presents its own challenges and requires additional refinement schemes. We therefore make use of further refinement prescriptions, outlined in \cite{2017BourneSijacki}. We briefly describe the relevant details of these schemes.

Firstly, from the jet axis out to a (cylindrical) radius of $\gamma r_{\rm Jet}$, the cell masses are forced below a spatially dependent maximum mass $m_{\rm cell}^{\rm max}(r)$ which ensures that the jet cylinder is sufficiently well populated. $m_{\rm cell}^{\rm max}(r)$ decreases as one gets closer to the jet axis according to   
\begin{equation}
    m_{\rm cell}^{\rm max}(r)= \Bigg[(\alpha - \beta)\bigg(\frac{r}{r_{\rm Jet}}\bigg)^\kappa + \beta\Bigg]M_{\rm cyl} \, ,
\end{equation}
where r is the (cylindrical) distance to the jet axis, $\alpha$, $\beta$ and $\gamma$ are parameters to be chosen and $\kappa$ is given by
\begin{equation}
    \kappa = \frac{1}{\ln(\gamma)}\ln\bigg(\frac{1-\beta}{\alpha - \beta}\bigg) \, ,
\end{equation}
meaning that
\begin{equation}
    m_{\rm cell}^{\rm max}\big(\gamma\, r_{\rm Jet}\big)  = M_{\rm cyl}\, .
\end{equation}
Both this refinement scheme and the black hole refinement scheme detailed above only act within the $R_{\rm ref}$ (which always encloses the jet cylinder) and so are localised to the vicinity of the black hole. Without further refinement schemes, cells into which jet energy, momentum and mass are injected are propelled out of this region and will then rapidly de-refine due to \textsc{arepo}'s mass refinement criterion \citep{2010Springel, 2020Weinberger+}. 

We identify jet material using a passive tracer which tracks the jet material, $m_{\rm i,J}$, in each cell. We set the tracer mass of cells in the jet cylinder (into which jet material is injected) to the cell mass, $m_i$, and then the tracer is advected with the gas. We make use of this `jet tracer' to ensure that there is sufficient resolution in the jet lobes by refining cells with a jet-fraction ($f_{i,{\rm J}}\equiv m_{i,{\rm J}}/m_i$) greater than $0.01$ if their volume satisfies
\begin{equation}
    V_i > \big[1- \log_{10}(f_{i,{\rm J}})\big]\,V_{\rm J}^{\rm max} \; ,
\end{equation}
where $V_{\rm J}^{\rm max}$ is a parameter to be chosen.

We also employ further jet refinement schemes to stop cells de-refining if the gradients between neighbouring cells are too steep \citep{2017BourneSijacki} and to limit the volume of cells as well as the gradients between neighbouring cells \citep{2017Weinberger+}.

\subsection{Timesteps}
The assumptions inherent in the \cite{1973ShakuraSunyaev} thin disc model and those in our simplified model of the Bardeen-Petterson effect introduce timescales that need to be resolved. We therefore ensure that the black hole timestep satisfies 
\begin{equation}
    \Delta t \leq 0.1 \,{\rm min}\bigg(\tau_{\rm GM},\, \frac{M_{\rm d}}{\dot{M}_{\rm d}} \bigg)\,,
\end{equation}
where $M_{\rm d} / \dot{M}_{\rm d}$ approximates the disc-draining time. Often, however, further constraints on the black hole timestep that are not specific to this sub-grid model force it to lower values (for further discussion see \cite{2018Fiacconi+}).

\section{Simulation setup}
\label{Sec: Setup}
Gas funnelled towards the centre of a galaxy due to secular evolution, large-scale inflows or mergers may settle into a rotationally supported disc surrounding the black hole. Indeed, such circumnuclear discs have been found to be ubiquitous in interacting galaxies and those with signatures of recent mergers \citep{2014Medling+, 2013ImanishiNakanishi, 1998DownesSolomon}. Observations have also shown that circumnuclear discs are preferentially found in Seyferts compared to gas-rich spirals with no nuclear activity \citep{2013Hicks+, 2012Garcia-BurilloCombes, 2007Davies+}. These circumnuclear discs have radii $\sim100$~pc and masses $10^{8-9}\; {\rm M_\odot}$ \citep{2007Chou+, 1999Schinnerer+}.
Motivated by the possibility of these discs residing at the centres of radio-loud Seyferts we simulate an isolated circumnuclear disc embedded within a stellar bulge and a single black hole at the centre. We surround the disc with warm, diffuse gas, with properties analogous to that found in the CGM which  provides a medium into which the jet can propagate.

Modelling just the circumnuclear disc and neglecting any effects of the surrounding galaxy, under the assumption that the circumnuclear disc is long lived and in equilibrium, provides an ideal setup to test our model due to its small scale nature as it allows us to focus our computational resources on the jet propagation. Furthermore, this allows rapid evaluation of the model and assessment of our parameter choices whilst still resolving the relevant spatial and time scales. 

\subsection{Initial conditions}
The initial placement of the mesh-generating points for the gas in the circumnuclear disc and collisionless particles in the stellar bulge are generated using a modified version of the procedure outlined in \cite{2005Springel+}.

The stellar bulge is modelled using a Hernquist density profile \citep{1990Hernquist}
\begin{equation}
    \rho_*(r) = \frac{M_*}{2\pi}\frac{r_{\rm s}}{r\,(r+r_{\rm s})^3}\, ,
\end{equation}
where the bulge mass is $M_* = 5\times 10^8 \; {\rm M_\odot}$ and the scale radius is $r_{\rm s} = 100 \; {\rm pc}$.

The gas disc is initialised with an exponential surface density profile
\begin{equation}
    \Sigma_{\rm gas}(R, z) = \frac{M_{\rm gas}}{2\pi \, h^2}\exp(-R/h) \, ,
\end{equation}
where $M_{\rm gas} = 10^8 \; {\rm M_\odot}$ is the disc mass and $h\approx 70 \; {\rm pc}$ is its scale-length. The initial vertical structure of the gas disc is determined such that it is in hydrostatic equilibrium and is initially isothermal with temperature $2\times10^4\; {\rm K}$.

We sample the stellar bulge with $2\times10^6$ collisionless particles of mass $m_* = 250 \; {\rm M_\odot}$ and with gravitational softening length $\epsilon_* = 5\; {\rm pc}$. The gas disc is initialised with $4\times 10^5$ mesh generating points with target mass $m_{\rm t} = 250 \; {\rm M_\odot}$ and gravitational softening length $\epsilon_{\rm gas} = 0.5\; {\rm pc}$. 

At the centre we place a black hole of mass $M_{\rm BH} = 10^6 \; {\rm M_\odot}$ and gravitational softening length $\epsilon_{\rm BH} = 5\; {\rm pc}$. This black hole mass is consistent with those found in late type galaxies with comparable stellar contents \citep{2019Greene+}.

We then add additional mesh generating points to fill the box with a diffuse, hot CGM. The gas cells are placed at rest, with temperature $10^7$~K and with their positions sampled from a uniform distribution to give a constant density. We consider the case where this CGM material has density $\rho_{\rm CGM} = 10^{-27} \; {\rm g\,cm^{-3}}$ (chosen such that the circumnuclear is in approximate pressure equilibrium with the CGM) which we refer to as the `standard' CGM and we also consider the case where the CGM is two orders of magnitude more dense, i.e. $\rho_{\rm CGM} = 10^{-25}\; {\rm g\,cm^{-3}}$, which we refer to as the `dense' CGM.

These initial conditions are then evolved for $200 \; {\rm Myrs}$ (approximately $4$ orbital periods of the disc) to allow transient features in the disc to dissipate and for the disc to come into pressure equilibrium with the CGM. The increased pressure in the `dense' CGM causes noticeable flattening of the circumnuclear disc and more instabilities are seen along the disc-CGM interface, however the disc is still able to reach quasi-steady state during the relaxation period.

\subsection{Additional tracers}
\label{Subsec: Tracers}
In these simulations we would like to be able to track how much disc material is physically entrained by the jet. When generating the initial conditions, we therefore initialise a passive Lagrangian tracer, $m_{i, {\rm disc}}$, in all gas cells in the circumnuclear disc, which we refer to as the `disc tracer'. The value of this disc tracer is then set to zero in all cells into which jet mass, momentum and energy are injected meaning that it tracks pure disc material.

Since our disc and jet tracers are independent, we obtain an effective CGM tracer for each cell via
\begin{equation}
    m_{i, {\rm CGM}} = m_{\rm i} - m_{i, {\rm J}} - m_{i, {\rm disc}} \, .
\end{equation}
We use this tracer to asses the levels of mixing and entrainment of CGM material in the jet lobes.

\subsection{Additional refinement}
\label{Subsec: Additional refinement}
As discussed in Section~\ref{Subsec: Refinement}, the mass refinement scheme in \textsc{arepo} maintains gas cells at a chosen target mass, which we choose to be $m_{\rm t}=250\;{\rm M_\odot}$ as this allows us to maintain sufficiently high resolution in the circumnuclear disc. Since the velocity of propagation of jet material is often supersonic, cells in the CGM are frequently unable to refine before the jet material reaches them which is particularly problematic if we maintain relatively coarse spatial resolution in the CGM. This can lead to increased numerical mixing at the jet-CGM interface as well as instabilities being washed out and the jet morphology and propagation being altered significantly \citep[see e.g.][]{2017BourneSijacki}.

Due to the density contrast between the circumnuclear disc and the CGM it is not necessarily the case that a global target mass $m_{\rm t} = 250\;{\rm M_\odot}$ will yield sufficiently high resolution in the CGM to mitigate these problems. In the `dense' CGM we find that we are, in fact, able to achieve sufficient resolution in the CGM with just a global target mass of $250\; {\rm M_\odot}$. In the `standard' CGM, however, this is not the case and we, therefore, implement an additional refinement scheme in any simulation that has a `standard' density CGM.

Since it is numerically infeasible to increase the resolution in the entire CGM, we enforce a spatially dependent, piecewise-defined target mass that is specifically applied to the gas cells in the CGM\footnote{Note that we do not apply this refinement scheme to cells in the jet. If the refinement scheme were only applied to cells in the disc then any such disc cell that enters the jet cylinder and becomes jet material will suddenly have its target mass decreased and we find that the resulting rapid refinement lead to artificial features in the jet.}, targeting the increased resolution to a central cylindrical region (with the cylinder axis parallel to the $z$-axis) in which the majority of the jet-CGM interaction is expected to occur. Specifically, we define the target mass for CGM cells at a cylindrical radius, $R$, to be
\begin{equation}
m_{\rm t}^{\rm CGM}(R) = 
\begin{cases}
    f_{\rm ref}\,m_{\rm t} & R < R^{\rm CGM}_{\rm ref}\\
    (a_{\rm ref} R + b_{\rm ref}) \, m_{\rm t} & R^{\rm CGM}_{\rm ref}<R<N_{\rm ref}\,R^{\rm CGM}_{\rm ref}\\
    m_{\rm t} & R > N_{\rm ref}\,R^{\rm CGM}_{\rm ref} 
\end{cases}
\end{equation}
where $f_{\rm ref}$, $N_{\rm ref}$ and $R^{\rm CGM}_{\rm ref}$ are free parameters.\\
The parameters $a_{\rm ref}$ and $b_{\rm ref}$ are chosen such that the $m_{\rm t}^{\rm CGM}(R)$ is continuous in $R$, i.e. 
\begin{align}
    &a_{\rm ref} = \frac{1}{R_{\rm ref}}\bigg(\frac{1 - f_{\rm ref}}{N_{\rm ref} - 1}\bigg) \, ,\\
    &b_{\rm ref} =\bigg(\frac{f_{\rm ref}\;N_{\rm ref}-1}{N_{\rm ref} - 1}\bigg) \, .
\end{align}

\section{Jets with fixed power and direction: Results}
\label{sec: fixed}
When implementing the sub-grid black hole accretion and Blandford-Znajek jet model, detailed in the previous sections, we first need to validate it against analytic expectations to verify the physical nature of our jets. Due to the complexity of the full sub-grid model, however, we first carry out a series of high-resolution studies in which the jet power and direction are fixed. Taking away these degrees of freedom makes for a more straightforward comparison with analytical models and creates a benchmark against which we can compare the results of runs that use our full sub-grid model. This simplified setup also allow us to more cleanly isolate how jets are expected to interact with the surrounding circumnuclear disc and the CGM.

In these idealised simulations we fix the jet power to a fraction of the Eddington luminosity of the black hole, $L_{\rm Edd}$\footnote{Here $L_{\rm Edd} \approx 1.26\times10^{44} \; {\rm erg\, s^{-1}}$ refers to the Eddington luminosity of the initial black hole mass, $M_{\rm BH} = 10^6 \; {\rm M_\odot}$, and disregards the fact that the Eddington luminosity of the black holes will change very slightly throughout the course of the simulation as the black hole accretes.}, and fix the direction of the jet to lie parallel to the z-axis.

Our simulation suite consists of six runs in total. We consider jets with three different powers: $10^{-4}L_{\rm Edd}$ - the `low' power run, $10^{-3}L_{\rm Edd}$ - the `medium' power run, and $10^{-2}L_{\rm Edd}$ - the `high' power run. These jets are then launched from the centre of the disc and we consider both the `standard' and `dense' CGM. We also fix the sub-grid velocity to $v_{\rm J} = 10^5\; {\rm km\, s^{-1}}$. Doing so removes all the spin-dependence of the jet energy, momentum and mass fluxes; these are constant and fully specified by equations~(\ref{eq: energy jet}) and (\ref{eq: momentum jet}). A full list of the parameters of these six simulations can be found in Table~\ref{tab: fixed}.

These runs are of a similar nature to those in \cite{2017BourneSijacki} in that they utilise the `kinetic' jet model and the jets have fixed direction. In our simulations, however, the jet power is fixed to some fraction of the {\it initial} Eddington luminosity of the black hole whereas the jet power in \cite{2017BourneSijacki} may change slightly as the black hole grows (note however that in those simulations the black hole growth was insignificant). In addition, we are probing significantly smaller scales\footnote{In \cite{2017BourneSijacki}, the target jet cylinder mass is $10^4 \;{\rm M_\odot}$ whereas in this work it is $0.5 \; {\rm M_\odot}$.} and, crucially, we are launching jets from a black hole embedded in a circumnuclear disc, whereas \cite{2017BourneSijacki} launched jets from black holes embedded in a hydrostatic atmosphere.

\begin{table*}
 \caption{Properties of the simulations in which the jet power and direction are fixed. The first and second columns show textual identifiers that we use throughout the results section when we refer to a specific simulation (e.g. the `high' power jet launched into the `dense' CGM). The power of the jet is shown in the third column. The fourth column corresponds to the density of the CGM. The fifth and sixth columns show the final time of each simulation and the length of the northern lobe of the jet at this time, respectively.}
 \label{tab: fixed}
 \begin{tabular}{lllcccc}
  \hline
  Power label & CGM label & Jet power & CGM density & Final time & Jet length at final time\\
  && [${\rm ergs \, s^{-1}}$ : $L_{\rm Edd}$]& [${\rm g \, cm^{-3}}$] & [Myrs] & [pc] \\
  \hline
  `high' & `standard' & $1.26\times10^{42}$ : $10^{-2}$ & $10^{-27}$ & $0.43$ & $2964$ \\
  `medium' & `standard' & $1.26\times10^{41}$ : $10^{-3}$ & $10^{-27}$ & $0.97$ & $2989$ \\
  `low' & `standard'& $1.26\times10^{40}$ : $10^{-4}$ & $10^{-27}$ &$2.54$ & $2995$ \\
  `high' & `dense'& $1.26\times10^{42}$ : $10^{-2}$ & $10^{-25}$ & $1.89$ & $2996$ \\
  `medium' & `dense' & $1.26\times10^{41}$ : $10^{-3}$ & $10^{-25}$ & $5.87$ & $2999$ \\
  `low' & `dense' & $1.26\times10^{40}$ : $10^{-4}$ & $10^{-25}$ & $10.00$ & $2120$ \\
  \hline
 \end{tabular}
\end{table*}

\begin{figure*}
    \centering
    \includegraphics[scale=1]{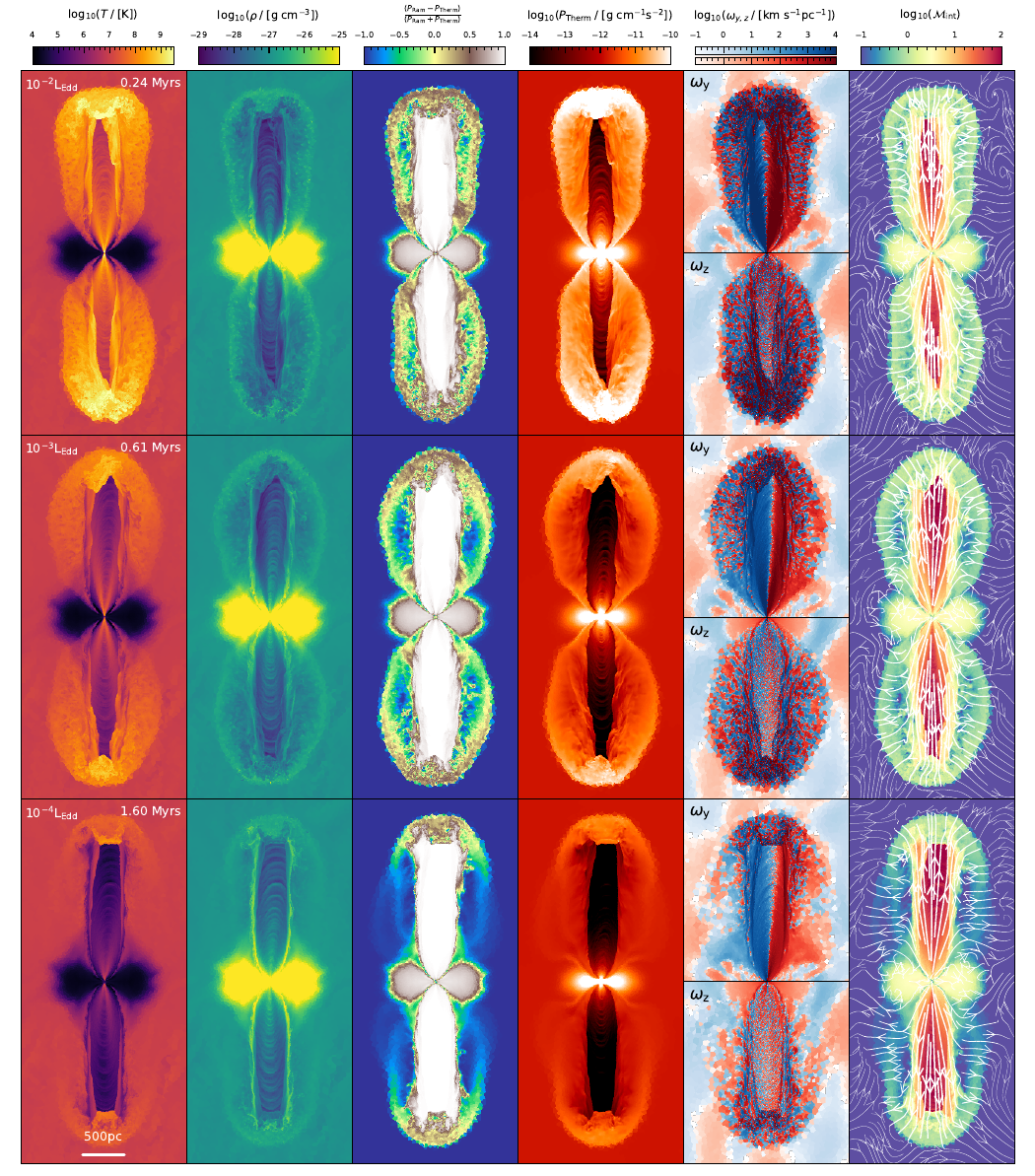}
    \caption{From top to bottom, each row shows slices through the computational domain for the `high', `medium' and `low' power jet runs in the `standard' CGM case. The first and second columns show the temperature and density fields, respectively. The third column shows a pressure diagnostic (which is described in Section~\ref{sec: Basic jet properties}). The fourth column shows the thermal pressure field and then the fifth column shows the $y$/$z$ component of the vorticity field in the top/bottom of each slice. Finally, in the sixth column, the slices indicate the internal Mach number of the gas, overlaid with streamlines of the velocity field in the $x$-$z$ plane, where the width of the streamline scales with the logarithm of the absolute velocity field. Each slice was created at the time at which the length of the northern lobe of the jet first exceeds $2000\; {\rm pc}$. This instant in time is then indicated in the top right-hand corner of the relevant temperature slice.}
    \label{fig: Slices thick fixedpower}
\end{figure*}
\begin{figure*}
    \centering
    \includegraphics[scale=1]{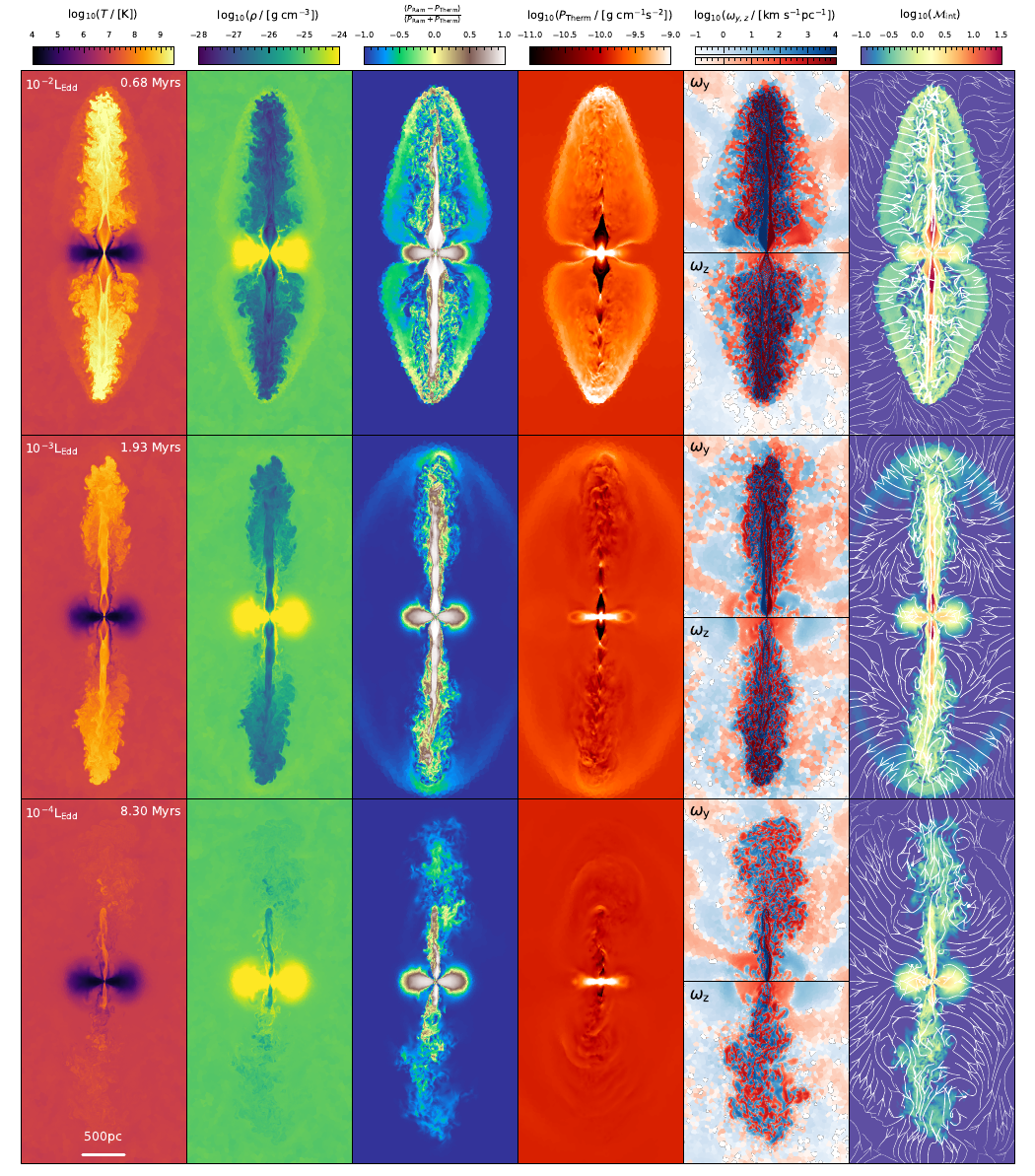}
    \caption{From top to bottom, each row shows slices through the computational domain for the `high', `medium' and `low' power jet runs in the `dense' CGM case. The first and second columns show the temperature and density fields, respectively. The third column shows a pressure diagnostic (which is described in Section~\ref{sec: Basic jet properties}). The fourth column shows the thermal pressure field and then the fifth column shows the $y$/$z$ component of the vorticity field in the top/bottom of each slice. Finally, in the sixth column, the slices indicate the internal Mach number of the gas, overlaid with streamlines of the velocity field in the $x$-$z$ plane, where the width of the streamline scales with the logarithm of the absolute velocity field. Each slice was created at the time at which the length of the northern lobe of the jet first exceeds $2000\; {\rm pc}$. This instant in time is then indicated in the top right-hand corner of the relevant temperature slice.}
    \label{fig: Slices thick dense fixedpower}
\end{figure*}

\subsection{Basic jet properties}
\label{sec: Basic jet properties}
We begin this section by providing a qualitative overview of the main morphological properties of these fixed power jets and discuss the physical processes that are driving the formation of these features. We perform this analysis with the aid of Figs.~\ref{fig: Slices thick fixedpower}~and~\ref{fig: Slices thick dense fixedpower} which show slices through the computational domain of various diagnostic quantities for the runs in the `standard' and `dense' CGM, respectively. From top to bottom, each row shows the relevant slices for the `high', `medium' and `low' power jet runs. Since the propagation timescales of the jets varies significantly from run-to-run and we wish to make comparisons between the jets on the same spatial scale, we select the snapshot at which the length of the northern lobe of each jet first exceeds $2000\; {\rm pc}$ and generate these slices at this time. The relevant time for each run is then indicated in the top right-hand corner of the corresponding temperature slice.

In each figure, the first and second columns show the temperature and density fields, respectively. To better identify the regions in which ram pressure or thermal pressure dominate, we define a pressure ratio
\begin{equation}
    P_{\rm ratio} = \frac{P_{\rm ram} - P_{\rm therm}}{P_{\rm ram} + P_{\rm therm}}\,,
\end{equation}
where $P_{\rm ram} = \rho u^2$ is the ram pressure and $P_{\rm therm}$ is the thermal pressure \citep{2017BourneSijacki}. Slices of this quantity are shown in the third column. In the fourth column we show the thermal pressure field and then in the fifth, we show the $y$/$z$ component of the vorticity field in the top/bottom of each slice, where the red and blue colour maps indicate regions where the relevant vorticity component has a positive or negative sign, respectively. Finally, in the sixth column, the slices indicate the internal Mach number of the gas, onto which we have overlaid streamlines of the velocity field in the $x$-$z$ plane, where the width of the streamline scales with the logarithm of the absolute velocity field.

Unsurprisingly, in all runs we can clearly identify the centrifugally supported circumnuclear disc from which the jet has been launched. In the `dense' runs, however, the disc has a noticeably smaller scale-height and scale-length as the higher pressure in this CGM has somewhat compressed the disc gas. Both in the `standard' and in the `dense' case, when the jets first become active they immediately encounter resistance from this cold, dense disc material and drive strong shocks into the disc. In general, the `high' power jets are more effective at pushing aside the disc material and emerge into the CGM with highly supersonic bulk velocities. The `low' power jets, however, are not as effective at overcoming the inertia of the disc material and, as a result, are colder and slower, albeit still supersonic. This can be quantified by considering the maximum Mach numbers within the jet material during the first $0.1\; {\rm Myrs}$ after the jet is launched. For the high power jet we find Mach numbers of $\sim270$ and $\sim160$ for the `standard' and `dense' jets, respectively, whereas those found in the `low' power jets are significantly lower: $\sim 50$ for the `standard' run and $\sim 40$ for the `dense' run, although these still indicate the presence of strong shocks. As the jets leave the disc, some of the disc material is entrained in the outflow, clearly identifiable in all runs by the dense regions surrounding the jet channels and, indeed, close to the base of the jet this disc component dominates the mass of the jets (see Sections~\ref{Subsec: Evolution of the jet lobes}~and~\ref{Sec: Distribution of energy and mass} below for further discussion).

In all runs, the launching of the jets also drives lateral shocks into the disc which advance in an almost hourglass shape in the radial direction, significantly altering the thermodynamic profiles of the inner disc, with the higher power jets able to influence disc material out to larger radii. These shocks rapidly weaken as they propagate outwards, eventually broadening into sound waves, leaving the outer disc largely unperturbed for the duration of these simulations. The interaction of these shocks and their reflections sets up a complex pattern of interference which mediates the properties of the gas in the vicinity of the black hole and as we do not allow the disc gas to radiatively cool these waves are not able to damp efficiently.

Outside of the disc, the launching of the jet drives a strong horizontal shock into the CGM with lateral velocity significantly higher than the velocity of the shock travelling within the disc material. This leads to an almost peanut-shaped shock front that advances into the CGM, particularly apparent in the pressure maps. In all runs this shock front initially envelops the whole jet, however, with time the strength of its horizontal advance diminishes and eventually becomes subsonic, leaving only the head of the jet driving shocks into the CGM. Over time these forward shocks also broaden into sound waves and detach from the head of the jet. At the times at which the slices in Figs.~\ref{fig: Slices thick fixedpower}~and~\ref{fig: Slices thick dense fixedpower} were made, the `medium' and `low' power jets in the `dense' CGM are no longer driving shocks into the CGM, whereas the `high' power jet in the `dense' CGM and all jets in the `standard' CGM are still mostly enveloped by shock fronts.

Turning now to the jets themselves, it is evident from inspection of Figs.~\ref{fig: Slices thick fixedpower} and \ref{fig: Slices thick dense fixedpower} that the morphology and propagation of these jets are sensitive both to the power of the jet and to the density of the surrounding medium. These figures also clearly show that, on the time and length scales considered here, the jets in the `dense' CGM are particularly sensitive to changes in jet power as they display a much more diverse range of morphological characteristics than those launched into the `standard' CGM which look comparatively similar.

In Fig.~\ref{fig: Slices thick dense fixedpower} the pressure and temperature slices highlight the series of recollimation shocks that have formed along the jet axis in all the `dense' CGM runs. These shocks form because, as the jets exit the disc, they are initially overpressured with respect to the external CGM and expand laterally until eventually they become under-pressured and are focused by the surrounding CGM. Since perturbations at the jet channel boundary are communicated by acoustic waves, the jet channel boundary oscillates as it repeatedly overshoots the equilibrium configuration, forming the recollimation shocks. Due to the lower CGM pressure in the `standard' CGM runs the length-scales over which these jets are recollimated are larger than the scales considered in Fig.~\ref{fig: Slices thick fixedpower}. We do, however, observe that the jet beams narrow towards terminal shock as the recollimation process begins. At later times, when these jets are eventually recollimated, they however do not display the repeated pattern of shocks that is so obvious in the `dense' CGM runs.

In the `dense' CGM runs the jet beam gas alternately heats and is compressed as it passes through the recollimation shocks and cools and expands as it passes through the rarefaction fans. Once the first recollimation shock has formed in each run, it remains approximately stationary with a spatial offset from the black hole that increases with jet power, ranging from $\sim 150 \; {\rm pc}$ in the `low' power run to $\sim 400 \; {\rm pc}$ in the `high' power run. The fact that the jets in the `standard' CGM are recollimated on significantly larger length-scales is indicative that, in this setup, it is the CGM density rather than the power of the jet that has a more significant influence on the timescale of the recollimation process.

Upstream of this first recollimation shock, we find fluid instabilities developing in the material surrounding the jet channel. However, it is downstream where these instabilities come to dominate the flow structure where they act to disrupt the jet beam and divert shocked jet material off-axis, enhancing the levels of turbulence and mixing in the surrounding cocoon. Indeed, as evidenced in Fig.~\ref{fig: Slices thick dense fixedpower}, in the `low' power jet these instabilities have completely disrupted the jet beam whereas the `high' power jet still retains a ram pressure-dominated beam which terminates in a weaker shock, close to the location of the forward shock being driven into the CGM.

As noted above, the morphology of the jets in the `standard' CGM runs is fairly similar, but there are some differences in the thermodynamic properties that are worth noting. Looking specifically at the jet channel region, as we move from low to high jet powers the temperature increases and the density decreases in general. The decrease in temperature with jet power can be used to understand why in Fig.~\ref{fig: Slices thick fixedpower} we find the highest internal Mach numbers in the `low' power jet. Since this quantity corresponds to the ratio of the absolute velocity of each gas cell to its sound speed, as the jet power decreases, the lower sound speeds associated with lower temperatures are sufficient to offset the decreasing velocities.

The temperature, density and pressure maps of the `standard' CGM simulations show evidence for a series of weak internal shocks that are propagating along the jet channel as shells of material travelling with different velocities collide. These velocity differences may be due to changes in the thermodynamic properties of the injection region which lead to fluctuations in the initial mass loading but we also expect that interactions with the circumnuclear disc play a significant role here. While we do find evidence for similar internal shocks in the jets in the `dense' CGM runs these are harder to see in Fig.~\ref{fig: Slices thick dense fixedpower} due to the smaller width of the jet channels.

The narrowing of the jet channel towards the head of the jets in the `standard' CGM runs leads to the formation of oblique shocks along the channel boundary and the jets then terminate in a strong annular shock at the pole of the jet axis in which the majority of the ram pressure dominated jet beam material undergoes thermalisation. This then leads to the formation of a `hot spot' at the stagnation point of the flow as the beam material is essentially brought to a halt relative to the downstream contact discontinuity. This hotspot, particularly visible in the pressure and temperature maps, is more obvious in the case of the `high' power jet. We do, however, see evidence for similar features in the `medium' and `low' power jets. The ram pressures of the beam and shocked CGM felt by jet material passing through the terminal shock causes it to be deflected laterally to regions of lower total pressure and it forms a wide fan of backflowing material. This material, along with that which has passed through the oblique shocks, feeds the low-density cocoon of shocked jet material. Conversely, in the `dense' CGM runs, the strongest on-axis shock is the first recollimation shock through which jet material passes and so the cocoon is fed due to the action of instabilities that have been excited by the recollimation shocks which act in addition to the terminal shock, where it exists.

For all three jet powers in the `dense' CGM case there is a significant enhancement in both components of the vorticity, particularly in the turbulent cocoon but also in the shocked CGM enveloped by the bow shock. There is also evidence for enhanced levels of vorticity in all of the `standard' runs, with greater levels of vorticity generation found for higher power jets. Vorticity enhancements are expected to come about due to turbulence and g-modes. The fact that we do not see evidence in any of the runs for significant vorticity enhancements outside of the jet cocoon indicates that these jets are not able to drive significant turbulence in the CGM \citep{2017BourneSijacki, 2017Weinberger+, 2016YangReynoldsA, 2015Reynolds+} and that any g-modes that have been excited are trapped within the cocoon for the duration of our simulations \citep{2015Reynolds+}.

\subsection{Comparison with analytic predictions}
\label{Subsec: Comparison with analytic predictions}
\begin{figure}
    \centering
    \includegraphics[width=0.5\textwidth]{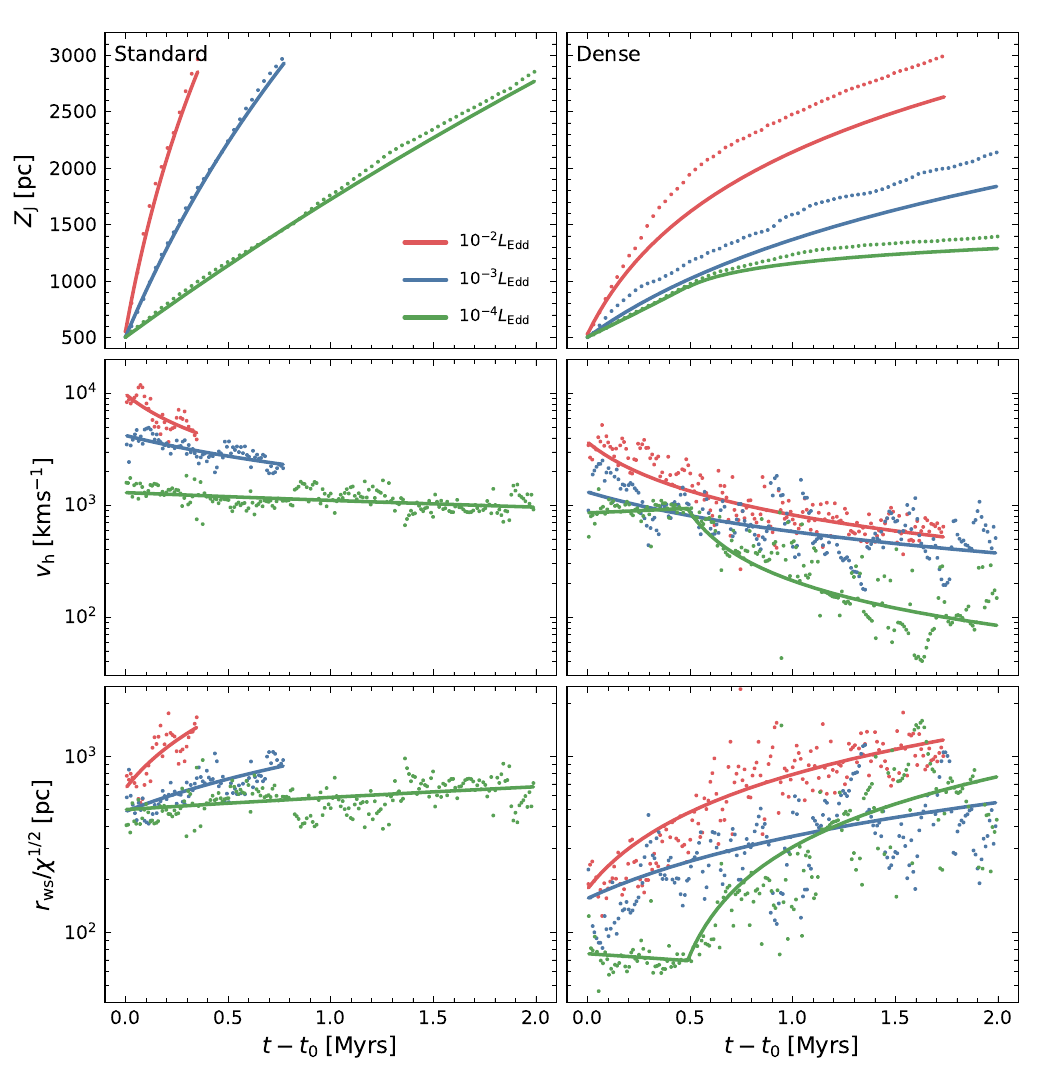}
    \caption{Individual points correspond to a value measured from the simulation data and the corresponding lines are the best-fit analytic model for the jet length (top row), the velocity of the jet head (middle row) and $r_{\rm ws} / \chi^{1/2}$ (bottom row). These are plotted as a function of $t-t_0$ where $t_0$ is the time at which the relevant jet has length: $Z_{\rm J, 0} = 500 \; {\rm pc}$. The `standard' CGM runs are shown in the left column and the `dense' CGM runs are shown in the right column.}
    \label{fig: Analytics}
\end{figure}

Having explored some of the salient features of these jets we now undertake a more quantitative analysis and compare their evolution to analytic predictions. We do not perform these comparisons with the aim of showing that our jets are well described by a specific model since we expect that the initial interaction with the circumnuclear disc will affect the long-term evolution of the jet (at least to some extent) and, as highlighted in Section~\ref{sec: Basic jet properties}, significant instabilities can develop in certain cases. We perform these comparisons, rather, with the aim of identifying regimes in which assumptions inherent to the analytic model either hold or break down which can be used to aid our understanding of the dominant processes affecting jet evolution.

Several works have developed analytic models that predict the time evolution of the jet length \citep{2018Harrison+, 2011Bromberg+, 2007PeruchoMarti, 1997KaiserAlexander, 1991Falle, 1989BegelmanCioffi}. Here, we choose to provide a comparison against the seminal analytic model detailed in \cite{1989BegelmanCioffi} which predicts the time evolution of the length of the jet by balancing the thrust of the jet, $\Pi_{\rm J}$, with the ram pressure force of a uniform density CGM\footnote{\cite{1989BegelmanCioffi} actually describe the medium into which the jet is launched as the ICM. Since the only relevant assumptions are that the medium has uniform density, we here refer to it as CGM material, to remain consistent with our setup.}:
\begin{equation}
\label{eq: ram pressure balance}
    \rho_{\rm CGM} v_{\rm h}^2 A_{\rm h} = \Pi_{\rm J} \, , 
\end{equation}
where $A_{\rm h}$ is the cross-sectional area of the bow shock at the end of the cocoon and $v_{\rm h}$ is the velocity of the head of the jet. \cite{1989BegelmanCioffi} assume that the thrust of the jet is given by $\Pi_{\rm J} \sim L_{\rm J} / v_{\rm J}$ where $L_{\rm J}$ is the jet power and $v_{\rm J}$ is the jet velocity.

\cite{2017BourneSijacki} used this model to validate the jets in their simulations, under the assumption that the jet thrust is given by $\Pi_{\rm J} =\chi\dot{M}_{\rm J}v_{\rm J}/2$ where $\dot{M}_{\rm J}$ is the sub-grid jet mass flux, $v_{\rm J}$ is now the sub-grid jet velocity and the factor of $1/2$ comes about due to the momentum flux being split between the two halves of the jet cylinder. The factor $\chi$ is a constant which takes into account any momentum boost that occurs in the initial launching of the jet. \cite{2017BourneSijacki} also expressed the cross-sectional area of the bow shock in terms of a working surface radius: $A_{\rm h} = \pi r_{\rm ws}^2$ and, under the above assumptions, found that their jets were a fairly good fit to the analytic model if the working surface radius scales linearly with time.

It should be noted here that implicit in the choice that $\chi$ be a constant is the assumption that the entire momentum flux at the base of the jet, $\chi \dot{M}_{\rm J}v_{\rm J}/2$, is what is felt at the head of the jet. In our analysis, we relax the assumption that $\chi$ is constant and instead of assuming that $r_{\rm ws}$ scales linearly with time, we make a more general assumption that $r_{\rm ws}/\chi^{1/2} = at+b$ for some constants $a$ and $b$. This is consistent with the fitting procedure in \cite{2017BourneSijacki} but alters the interpretation of the result. In our case, it should be made clear that $r_{\rm ws} / \chi^{1/2}$ does not give us any information about the behaviour of the working surface radius or the momentum flux at the head of the jet as the two are degenerate.

To determine the analytic jet length evolution we obtain the velocity of the jet head from equation~(\ref{eq: ram pressure balance})
\begin{equation}
\label{eq: vh}
    v_{\rm h} \approx \bigg(\frac{\chi\dot{M}_{\rm J}v_{\rm J}}{2\pi\rho_{\rm CGM}}\bigg)^{1/2}\frac{1}{r_{\rm ws}} \, .
\end{equation}
Inserting $r_{\rm ws}/\chi^{1/2} = at+b$ then gives
\begin{equation}
    v_{\rm h}(t) \approx \bigg(\frac{\dot{M}_{\rm J}v_{\rm J}}{2\pi\rho_{\rm CGM}}\bigg)^{1/2}\frac{1}{at+b} \, ,
\end{equation}
and integrating this then gives the jet length
\begin{equation}
    Z_{\rm J}(t) \approx \bigg(\frac{\dot{M}_{\rm J}v_{\rm J}}{2\pi\rho_{\rm CGM}}\bigg)^{1/2}\frac{1}{a}\ln\bigg(\frac{at + b}{at_0+b}\bigg) + Z_{\rm J,0}\, ,
\end{equation}
where $t_0$ is the time at which the jet length is $Z_{\rm J,0}$.\\

Since initial interactions with the circumnuclear disc (and differences in the initial mass-loading between the north and south halves of the jet cylinder) can lead to slight differences in the evolution of the north and south lobes of a given jet (which are typically more pronounced at early times), we restrict our analysis to the northern jet lobe. Further to this, since we know this model is not valid for the initial evolution of the jets before they are able to break out of the circumnuclear disc, for each jet we begin the comparison at the time when its length has reached $Z_{\rm J,0} = 500 \; {\rm pc}$. We then follow the evolution of each jet for a period of time that is sufficient to be able to make robust comparisons with the analytic model (specifically, until its length reaches $3000 \; {\rm pc}$ or $2\; {\rm Myrs}$ has elapsed). After the length of the northern jet lobe has been calculated, the velocity of the jet head is found by differencing the jet length between snapshots and we then fit for the constants $a$ and $b$. 

\begin{figure*}
    \centering
    \includegraphics[]{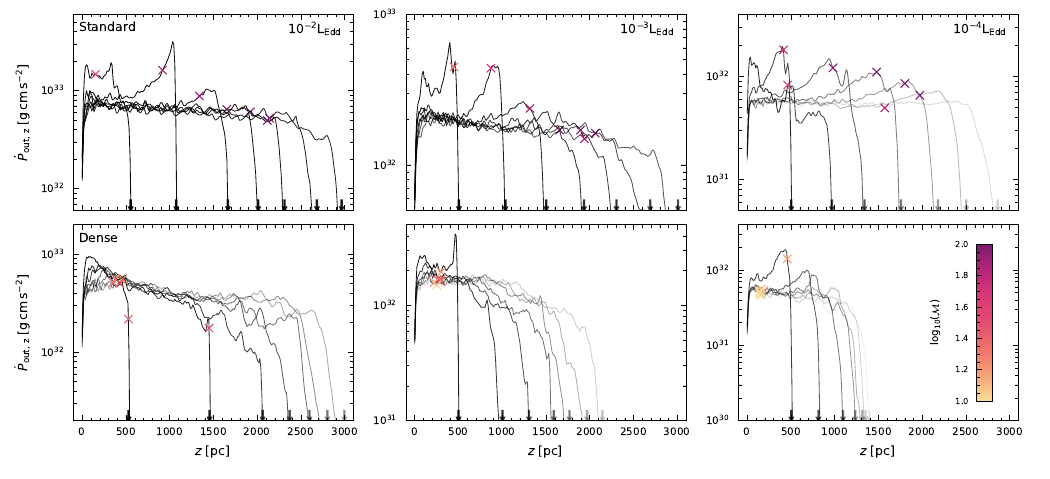}
    \caption{The positive flux of $z$-momentum in jet material as a function of distance along the jet axis. The top and bottom rows show results from runs in the `standard' and `dense' CGM cases, respectively. Each column, from left to right, corresponds to the `high', `medium' and `low' power jets. Individual lines show the momentum flux profile for the relevant jet at a specific time where the times have been chosen to sample the whole range for which the analytic fitting procedure was carried out (see Fig.~\ref{fig: Analytics}). Lighter lines correspond to a larger value of $t-t_0$. For each line, the arrow on the $x$-axis with the corresponding weight indicates the length of the jet.}
    \label{fig: Momentum flux}
\end{figure*}

The values measured from the simulation and the best fit analytic models are displayed in Fig.~\ref{fig: Analytics}. The top row shows the jet length, the second row shows the velocity of the jet head and the third row shows the resulting value for the quantity $r_{\rm ws} / \chi^{1/2}$. Each individual point corresponds a value measured from the simulation data and the solid line shows the best-fit analytic prediction. The results for the jets in the `standard' CGM case are shown in the left column and the jets in the `dense' CGM case are shown on the right\footnote{Whilst the fits were performed for each jet using the data from all snapshots in the relevant time interval, for ease of viewing we have plotted every third data point for the jet lengths.}.

Inspecting this figure it is immediately clear that, on the whole, the analytic model provides a relatively good fit to the jets in the `standard' CGM. These runs also display the expected scaling of the jet length and velocity with jet power, with the largest jet head velocities found in the `high' power jet. The jets in the `dense' CGM, however, do not seem to be well described by this model and show significantly more variability in the measured velocity. Moreover, the `low' power jet in the `dense' CGM shows two distinct phases in its evolution with the initial jet propagation stalling around $1.17\; {\rm Myrs}$ after the jet turns on (which corresponds to $t - t_0 = 0.49\; {\rm Myrs}$) after which its length stagnates. This behaviour is obviously not well fit by just a single pair of values, ($a$, $b$), and so the analytic fit we provide in Fig.~\ref{fig: Analytics} corresponds to a piece-wise function which better reflects this bimodal behaviour. Whilst there is not such a clear transition in the `medium' and `high' power jet runs, there are still differences between their early and late time behaviour that have not been well captured by the analytic model.

We further examine this by considering the positive flux of z-momentum in jet material\footnote{Here we define jet material to be all cells with $f_{\rm J} > 10^{-3}$ (see Section~\ref{Subsec: Evolution of the jet lobes} for further discussion). Whilst this definition will lead to cocoon material that is outflowing being included in the calculation, we do not expect this to change our interpretation of the momentum fluxes.} as a function of distance along the jet axis, shown in Fig.~\ref{fig: Momentum flux}. The first and second rows show the results from runs in the `standard' and `dense' CGM cases, respectively, with each column corresponding to the `high', `medium' and `low' power jets from left to right. Each line then corresponds to the momentum flux profile for the relevant jet at a specific time where the times have been chosen to sample the whole range for which the analytic fitting procedure was carried out, with lighter lines corresponding to a larger values of $t-t_0$.

We have made use of \textsc{arepo}'s shock finder \citep{2015SchaalSpringel} to characterise the shocks within the jet material. On each line, the marker `x' shows the location of the strongest shock in the outflowing jet material at this time and the colour of the marker indicates the Mach number of this shock. As expected, following the discussions in Section~\ref{sec: Basic jet properties}, the strongest shock in the `standard' CGM runs is the reverse shock at the head of the jet. In the `dense' CGM runs, before the recollimation shock forms, the strongest shock is also the reverse shock at the head, but once the recollimation shock forms, this first shock through which jet material passes is strongest. Comparing the strength of the shocks we see that those in the `dense' CGM runs typically have lower Mach numbers than those in the `standard' CGM.  

For the `high' and `medium' power runs we see that, in general, the momentum flux decreases along the length of the jet, but this decrease is much more significant in the `dense' CGM runs and it primarily occurs beyond the recollimation shock, as material is thermalised and diverted off-axis.

We also see that for most jets that there is a peak in momentum flux at early times which propagates towards the head before settling down somewhat. This behaviour is akin to a transient in the initial conditions and occurs primarily due to the gas in the vicinity of the black hole readjusting to the launching of the jet. Specifically, as the density in the centre drops, the action of the black hole refinement allows the jet cylinder mass to fall down to its target mass which, in turn, increases the velocity of individual cells kicked out of the jet cylinder. The communication of these higher velocities to the head of the jet is what leads to the travelling peak seen in the momentum flux profiles. At higher jet powers, regulation of the gas properties in the vicinity of the black hole occurs on shorter timescales and so this feature persists for longer in the `low' power jets.

Crucially however, in all runs but particularly those in the `dense' CGM, the momentum flux at the head of the jet is not constant in time; rather it seems to decrease. It is therefore likely that the break down of the analytic model can be attributed to the assumption of jet momentum flux conservation and, in particular, in the `dense' CGM case this occurs due to the presence of strong recollimation shocks which thermalise jet material and the ensuing onset of significant instabilities which divert jet beam material off-axis. Furthermore, the fact that the momentum flux at the head of the jet is time-variable justifies our interpretation that $\chi$ should be characterised as a time dependent quantity that corresponds to the momentum flux felt by the head of the jet in units of the sub-grid momentum flux. In the case where multiple shocks and turbulence are acting, this is a complex non-linear function that is not well described by a simple analytical model. Since we do not expect the radius of the working surface to be well described by analytical arguments, we have no reason to expect that the quantity $r_{\rm ws} / \chi^{1/2}$, displayed in the bottom row of Fig.~\ref{fig: Analytics}, should be linear.

\subsection{Evolution of the jet lobes}
\label{Subsec: Evolution of the jet lobes}
We now turn to more quantitative analysis of the energy and mass content of the jet lobes to better understand the processes driving the morphological features we have identified. 

Throughout this discussion, we will restrict our attention to the `jet lobes' which we define to be all gas cells that have a jet tracer fraction satisfying:
\begin{equation}
    f_{\rm J, i} \equiv \frac{m_{\rm J, i}}{m_{\rm i}} > f_{\rm J}^{\rm (thresh)} \; ,
\end{equation}
where we choose the threshold jet fraction to be $f_{\rm J}^{\rm (thresh)} = 10^{-3}$  \citep{2017Weinberger+, 2013HardcastleKrause}. It should be noted, however, that this threshold jet fraction is not an empirically measurable quantity and, intuitively, we would expect it to be a complex function of jet power and time, rather than simply a constant. This work, however, is concerned with better understanding the underlying physical processes driving the morphology and evolution of jets rather than the precise determination of the mass and energy content of the jet lobes and we find that the choice of $f_{\rm J}^{\rm (thresh)} = 10^{-3}$ is sufficient for these purposes.

\begin{figure*}
    \centering
    \includegraphics[]{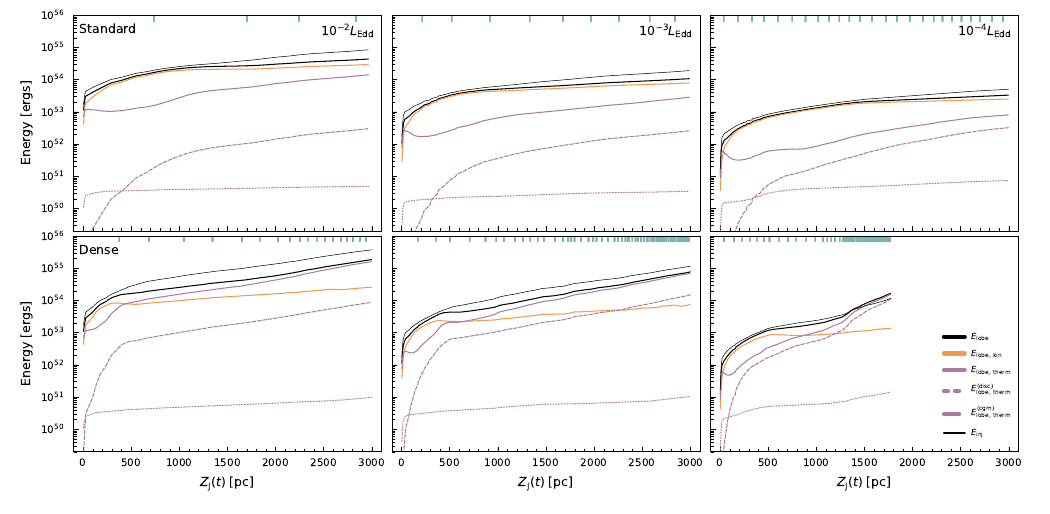}
    \caption{The energy composition of the northern lobe of each jet as a function of the total length of this lobe. The top and bottom rows show the runs in the `standard' and `dense' CGM cases, respectively, and from left to right the columns correspond to the `high', `medium' and `low' power jets. In each panel, the total energy contained within lobe material and the total energy that has been injected are shown by the thick and thin black lines, respectively.  The orange and purple lines show the kinetic and thermal components of this total jet lobe energy. Estimates to the thermal energy coming from CGM and disc material are shown by the dashed and dotted purple lines, respectively. To aid intuition as to the correspondence between the relevant jet length and elapsed time, on the top $x$-axis of each panel we have plotted blue ticks that are equally spaced in time.}
    \label{fig: jet energy evolution length}
\end{figure*}

\begin{figure*}
    \centering
    \includegraphics[]{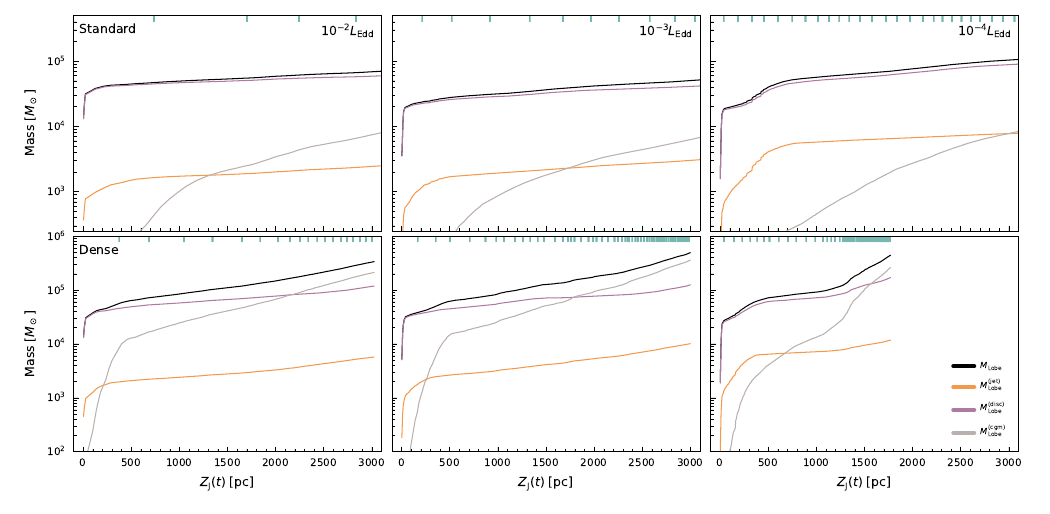}
    \caption{The mass of the northern lobe of each jet as a function of the total length of this lobe. The top and bottom rows show the runs in the `standard' and `dense' CGM cases, respectively, and from left to right the columns correspond to the `high', `medium' and `low' power jets. In each panel, the total mass of lobe material is shown by the thick black lines. The orange, purple and grey lines show the the jet, disc and CGM contributions to this total lobe mass. To aid intuition as to the correspondence between the relevant jet length and elapsed time, on the top $x$-axis of each panel we have plotted blue ticks that are equally spaced in time.}
    \label{fig: jet mass evolution length}
\end{figure*}

We begin this analysis by studying the evolution of the total energy and mass content of the jet lobes, shown in Figs.~\ref{fig: jet energy evolution length}~and~\ref{fig: jet mass evolution length}. As discussed in previous sections, the jet propagation timescale depends sensitively on the power of the jet as well as the density of the CGM. In order to make meaningful comparisons between all jet runs, it is therefore instructive to consider the evolution of the lobes as a function of the total length of the jet lobes {\it which acts as a proxy for time}, rather than as a function of time itself. To aid intuition as to the correspondence between jet length and the elapsed time, on the top $x$-axis of each panel we have plotted blue ticks that are equally spaced in time. In each plot, the `standard' CGM runs are shown in the top row and the `dense' CGM runs are shown in the bottom row. Each column corresponds to the `high', `medium' and `low' power jets, from left to right, respectively\footnote{Were we to perform the following analysis using data from both the northern and southern lobes, slight differences in their length evolution and the location of internal shocks can complicate things. We therefore restrict the analysis in this section and Section~\ref{Sec: Distribution of energy and mass} to the northern lobe of each jet and henceforth any reference to the 'jet length' refers to the length of the northern lobe of the relevant jet and similarly any quantitative reference to masses and energies are those of the northern lobe.}.

\subsubsection{Lobe energy evolution}
\label{Subsubsec: Lobe energy evolution}
In Fig.~\ref{fig: jet energy evolution length}, the total energy contained within the jet lobe and the total energy that has been injected into the northern lobe of the jet are shown by the thick and thin black lines, respectively, while the kinetic and thermal components of this total lobe energy are shown by the orange and purple lines, respectively\footnote{We wish to emphasise that following analysis should not be interpreted as predictions for the final partitioning of the injected jet energy since the volume we define to be the jet lobes can have net inflow/outflow. Similarly, once turbulence takes hold and disrupts the jet, our jet tracer definition of `lobes' will not refer to a coherent volume and so this analysis is better thought of as quantifying the energy content of jet-dominated material.}.

We can see that, in most runs, the total energy in the jet lobes remains below the injected energy throughout their evolution, as would be expected since the jet inflation does work on the surrounding CGM (and this measure of total energy does not account for any energy in jet material that does not satisfy our lobe criterion, although we expect this to have a negligible effect). In the `standard' CGM runs, the ratio of the lobe energy to the injected energy decreases with jet power, while in the `dense' CGM runs there is not a clear trend.

The jets in the `standard' CGM case all show similar behaviour: the kinetic energy component remains dominant throughout (with the average kinetic fraction ranging from $0.75$ - $0.85$), the only exception at very early times when the thermal content of the lobes is high due to initial shock-heating that occurs as the jet breaks out of the circumnuclear disc. At later times, even though the thermal energy content never becomes comparable to the kinetic, as the jets propagate further from the black hole the thermal component is more significant. The fraction of the lobe energy in thermal energy is larger for higher power jets and by the time the jet length reaches $3000\; {\rm pc}$ this amounts to $\sim 0.32$ for the `high' power jet, $\sim 0.26$ for the `medium' power jet and $\sim 0.24$ for the `low' power jet. This increase in thermal fraction correlates with the strength of the reverse shock at the head of the jet (see Fig.~\ref{fig: Momentum flux}), however other factors such as entrainment and mixing in the lobes will also be at play here.

Turning now to the runs in the `dense' CGM case, we see that they too are all dominated by thermal energy during the disc break-out phase and then quickly transition to kinetic dominance. In contrast to the runs in the `standard' CGM, however, the `dense' CGM runs all rapidly become thermally dominated again. The time and length-scales over which the transition to thermal dominance occurs decreases with increasing jet power, ranging from $1.08\; {\rm Myrs}$ ($Z_{\rm J} = 874\; {\rm pc}$) in the `low' jet power case to $0.13\; {\rm Myrs}$ ($Z_{\rm J} = 462\; {\rm pc}$) in the `high' jet power case.

To better understand the behaviour of the thermal energy of the jet lobes we consider the impact of two distinct processes that primarily influence this thermal content. Firstly, any mixing or entrainment that occurs, such as in the backflowing cocoon where shocked jet material mixes with hot CGM, can lead to an increase in thermal energy. Similarly, instabilities that develop along the jet beam will also lead to enhanced mixing, as will the entrainment of disc material. Note that as a result of mixing and entrainment the thermal energy of the lobes will increase, as will the mass of jet lobes (see Section~\ref{Subsubsec: Lobe mass evolution} below), but only mixed material that satisfies our jet tracer fraction threshold will contribute to the lobe thermal energy in Fig.~\ref{fig: jet energy evolution length}. The second channel through which changes to the energy content of the lobes result is shock heating. This will occur as jet material passes through on-axis shocks but also in the initial shocks as the jet breaks out of the disc and, where it exists, in the supersonic turbulence in the lobes.

To better understand how the thermalisation is being driven, for each run we have provided an estimate (which is likely a lower bound) on the thermal energy content of the jet lobes that we expect has come from entrained CGM material, which is indicated by the dashed purple line in Fig.~\ref{fig: jet energy evolution length}. This quantity has been calculated using our CGM tracer, under the assumption that the mass associated with this tracer has a specific internal energy equal to the initial specific internal energy of CGM material (which was initialised with temperature $T_{\rm CGM} = 10^7\;{\rm K}$). We have also carried out a similar analysis using the disc tracer to estimate the contribution of entrained disc material to the thermal energy of the lobes, under the assumption that disc material has temperature $T_{\rm disc} = 2\times 10^4\; {\rm K}$ (again, this is likely a lower bound as we find a non-negligible amount of disc tracer in the shock-heated cocoon material and some disc material will have undergone shock heating during the initial relaxation period). This estimate is shown by the dotted purple line in Fig.~\ref{fig: jet energy evolution length}. These estimates serve two purposes. Firstly, we can use them to make predictions about the relative importance of the thermal energy associated with disc and CGM material in the jet lobes. Secondly, since they are calculated under the assumption that the material has not undergone shock-heating we can use them to identify regimes where shock-heating must be important\footnote{It should be noted, however, that there is degeneracy between these two scenarios (for example, significant shock-heating of disc material may change its relative importance). Further, we should point out that it is not possible to perform similar analysis for the case of `pure' jet material as this would require us to associate with it a specific internal energy.}. 

We first examine our estimate of thermal energy associated with disc material. Looking specifically at the `high' power jet in the `dense' CGM case we find that our disc thermal energy estimate is, on average, $\sim 4$ orders of magnitude lower than the total thermal energy. In order to explain the total thermal energy, the disc material would therefore need to have be shock heated to $\sim10^8\;{\rm K}$. Whilst a non-negligible amount of disc material ends up in the cocoon in which we do find temperatures that exceed $10^8\; {\rm K}$, the majority of the disc material can be found in the cold dense sheath surrounding the jet channel (see Fig.~\ref{fig: Slices thick dense fixedpower}). For this reason it is unlikely that the entire thermal energy can be accounted for by entrainment of disc material, regardless of whether it has undergone shock-heating or not. Similar arguments can be made for the other runs. 

Turning now to our estimate of the contribution to the thermal energy from CGM material we find that (except at early times) it is much higher in comparison to our estimate of the disc thermal energy. In both the `standard' and `dense' CGM cases for lower jet powers, the thermal contribution from the CGM material becomes significant when compared to the total thermal energy of the lobes. At the final time we find, for these `low' power jets, our CGM thermal energy estimate is $0.40$ and $0.71$ times the total thermal energy in the `standard' and `dense' cases, respectively. Since our estimates are calculated under the assumption that the entrained CGM material has not undergone shock-heating we can infer that, for the thermal energy content to be dominated solely by that of the entrained CGM material, we must invoke significant heating in the `high' power cases whereas at lower powers, less heating would be needed. 

The effects of CGM entrainment are particularly evident in the case of the `low' power jet in the `dense' CGM. When this jet transitions to a state of thermal energy dominance, the thermal energy of the lobe comes predominantly from `unshocked' CGM material. This transition occurs as the jet beam starts to disrupt which is accompanied by the jet length stagnating (as evidenced by the blue ticks on the $x$-axis). Interestingly, we find that this ultimately leads to the energy in the lobe of this jet exceeding the total injected energy.

\subsubsection{Lobe mass evolution}
\label{Subsubsec: Lobe mass evolution}

We now examine the evolution of the mass of the jet lobes in each of our simulations, shown in Fig.~\ref{fig: jet mass evolution length}. Making use of our passive tracers (see Section~\ref{Subsec: Tracers}), we have further broken this lobe mass down into that coming from pure jet, entrained disc and entrained CGM material.

Considering the runs in the `standard' CGM case, it is apparent that the mass of these jets is entirely dominated by entrained disc material. The mass associated with `pure' jet material is subdominant at all times (which is unsurprising given the very large difference in typical densities of these two components). We find that the mass associated with CGM material is negligible in the initial stages of evolution, but increases as the jet propagates further from the black hole when processes such as mixing and entrainment become effective and backflows develop. At all times, however, the CGM lobe mass content is at least a factor of $4$ smaller than the contribution from disc material. Comparing the three different jet powers, it can be seen that the evolution of the mass of each jet is qualitatively similar, however, with increasing jet power, the time at which the mass in CGM material exceeds that coming from `pure' jet material decreases and occurs over shorter length-scales. This is consistent with what is seen in Fig.~\ref{fig: Slices thick fixedpower}, specifically that the levels of mixing and turbulence within the cocoon increase with jet power.

Turning now to the jets in the `dense' CGM case, the increase in the total mass of jet lobes is more significant than in the `standard' CGM runs, particularly at later times when the jet propagation begins to stall in the `low' and `medium' power jets. This gain in mass is primarily driven by a considerable increase in the mass of CGM material in the lobes which, while initially subdominant relative to disc material, becomes the dominant contributor to the lobe mass. This effect is much more pronounced in the `dense' CGM runs as mixing becomes more effective due to the development of recollimation shocks and fluid instabilities; in fact the final percentage contribution to the lobe mass from CGM material lies in the range $59$-$73$ per cent in the `dense' CGM runs, whereas in the `standard' CGM runs this same quantity lies in the range $14$-$19$ per cent.

One final point to discuss in Fig.~\ref{fig: jet mass evolution length} is the fact that the mass of jet material (material that has, at some point, been in the jet cylinder) in the lobes is higher in the `low' power jets. We would expect that, to first order, the total mass of jet material in the simulation scales linearly with time. Since the plots in Fig.~\ref{fig: jet mass evolution length} show the mass evolution as a function of jet length it is, therefore, expected that we would find a higher mass in the slower `low' power jets. The amount of pure jet material in the simulation will also depend on the mass of the jet cylinder and, as mentioned in Section~\ref{Subsec: Comparison with analytic predictions}, the cylinder is initially slightly over-massive until the gas flows in the vicinity of the black hole are able to be readjust to the jet which takes longer in the `low' power case. This could lead to more massive lobes in the `low' power jets, although we expect this effect to be relatively small.

\subsection{Distribution of energy and mass within the jet lobes}
\label{Sec: Distribution of energy and mass}
\begin{figure*}
    \centering
    \includegraphics[]{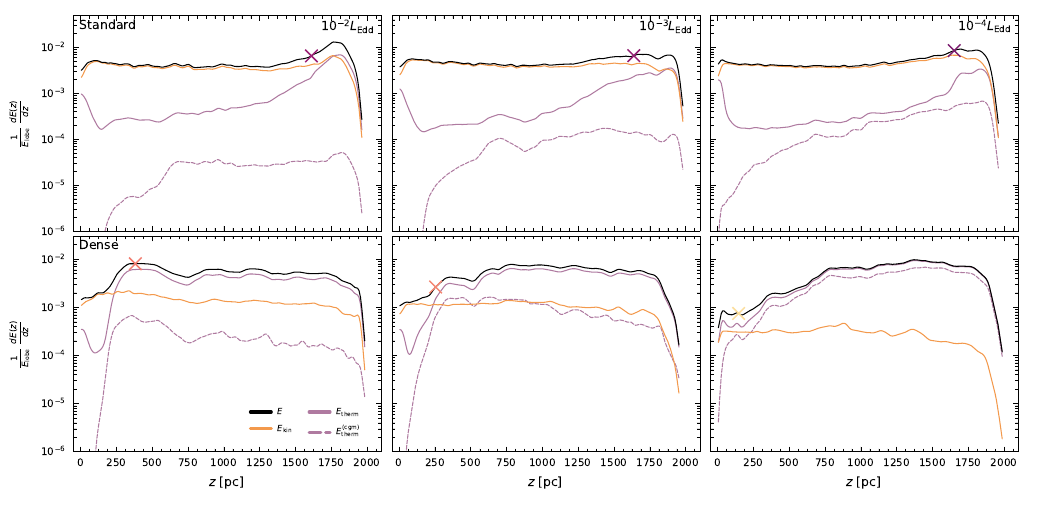}
    \caption{The normalised energy distribution of jet lobe material as a function of distance along the jet axis. In each plot, the runs in the `standard' CGM are shown in the top row and those in the `dense' CGM are shown in the bottom row. From left to right, each column corresponds to the `high', `medium' and `low' power jets. The profile for each jet was created at the same time as the slices in Figs.~\ref{fig: Slices thick fixedpower}~and~\ref{fig: Slices thick dense fixedpower} meaning that the length of the northern lobes of the jets are all $\sim2\;{\rm kpc}$. The black line in each panel shows the total energy profile. The location of the strongest shock within the jet lobe material is indicated by an 'x', where the colour of the marker encodes the Mach number of the shock, using the same colour-scale as in Fig.~\ref{fig: Momentum flux}. The kinetic and thermal components of this total energy are indicated by the orange and purple lines, respectively and the purple dashed line corresponds to our estimate of the thermal energy from the entrained CGM material.}
    \label{fig: Jet energy pdf}
\end{figure*}

\begin{figure*}
    \centering
    \includegraphics[]{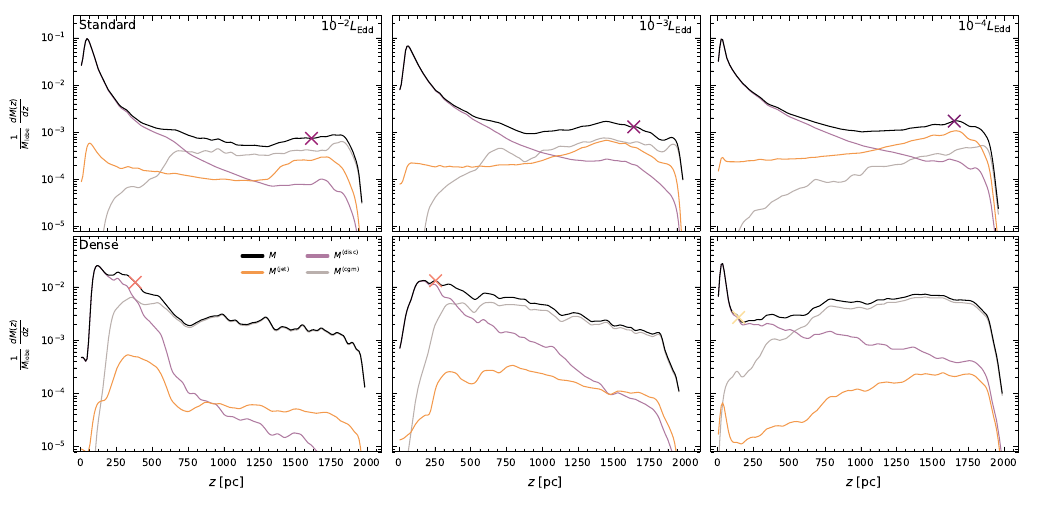}
    \caption{The normalised mass distribution of jet lobe material as a function of distance along the jet axis. In each plot, the runs in the `standard' CGM are shown in the top row and those in the `dense' CGM are shown in the bottom row. From left to right, each column corresponds to the `high', `medium' and `low' power jets. The profile for each jet was created at the same time as the slices in Figs.~\ref{fig: Slices thick fixedpower}~and~\ref{fig: Slices thick dense fixedpower} meaning that the length of the northern lobes of the jets are all $\sim2\;{\rm kpc}$. The black line in each panel shows the total mass profile. The location of the strongest shock within the jet lobe material is indicated by an 'x', where the colour of the marker encodes the Mach number of the shock, using the same colour-scale as in Fig.~\ref{fig: Momentum flux}. The jet, disc and CGM components of this total energy are indicated by the orange, purple and grey lines, respectively.}
    \label{fig: Jet mass pdf}
\end{figure*}

Having explored the evolution of the energy and mass content of the jet lobes, we now look more closely at the distribution of this energy and mass within the jet lobes {\it at a fixed point in time} during evolution of the jet. Specifically, Figs.~\ref{fig: Jet energy pdf}~and~\ref{fig: Jet mass pdf} show the normalised energy and mass distributions of the jet lobe material as a function of distance along the jet axis. We have marked the position of the strongest shock within the jet lobe material with an `x', where the colour of the marker encodes the Mach number of the shock, using the same colour-scale as in Fig.~\ref{fig: Momentum flux}. As anticipated, we find that in the `standard' CGM runs, this corresponds to the reverse shock at the head of the jet whereas in the `dense' CGM runs this locates the first recollimation shock. The profile for each jet was created at the same time as the slices in Figs.~\ref{fig: Slices thick fixedpower}~and~\ref{fig: Slices thick dense fixedpower}.
\subsubsection{Energy distribution}
\label{Subsubsec: Energy distribution}

In Fig.~\ref{fig: Jet energy pdf}, the total energy distribution is then further broken down into the kinetic energy, total thermal energy and thermal energy from entrained CGM material\footnote{ We have not included our estimate of the thermal energy from entrained disc material in Fig.~\ref{fig: Jet energy pdf} as, in all cases, it was clearly subdominant along the entire length of the jet.}.

First, considering the jets in the `standard' CGM runs, we see that the total jet energy is distributed largely uniformly along the jet out to the reverse shock and that the jets are kinetically dominated. Only close to the base of the jet is the thermal energy content comparable to that of the kinetic energy which is due to the shock-heating in the disc. Downstream of the reverse shock, the distribution of total energy in the lobe peaks in the `high' power jet and we observe an associated peak in the thermal energy distribution, driven by the thermalisation of material passing through the reverse shock. Similar behaviour is seen in the `medium' and `low' power jets but the peaks in the total energy distribution are not as prominent. It should be noted here that using a jet tracer fraction of $10^{-3}$ to define jet lobe material means that the majority of the cocoon, including backflowing material, is categorised as lobe material. The kinetic energy of this backflowing material will therefore contribute to the profiles in Fig.~\ref{fig: Jet energy pdf} and explains why in the `high' power jet run, the kinetic energy profile seems to peak downstream of the strong reverse shock.

The thermal energy of entrained CGM appears largely subdominant for the `high' and `medium' power jet. In the 'low' power jet, however, the thermal energy from CGM material does becomes comparable to the total thermal energy of the jet lobes. Recall however that our estimate of the CGM thermal energy contribution assumes that this material has not undergone heating. The observation that, in the `high' power jet, our estimate is significantly lower than that of the total thermal energy {\it does not mean} that the thermal energy of CGM material is not important here. Rather, it means that if it is to be important, then it must have undergone shock heating which is indeed more significant for higher power jets (see Section~\ref{Subsec: Evolution of the jet lobes}). 

Turning now to the jets in the `dense' CGM case, we see that as we move along the jet axis, there is a significant redistribution in the total energy content, driven by the increasing fraction of energy in thermal form. The thermal energy then remains dominant out to the head of the jet. The `high' power jet run shows a significant increase in the contribution of thermal energy to the total which peaks around the location of the first recollimation shock and remains approximately constant out to the head of the jet. Naively, it may be tempting to attribute this increase in thermal energy distribution to the action of the recollimation shock alone. Whilst material in the jet channel that passes through this shock will undergo thermalisation, the slices in Fig.~\ref{fig: Slices thick dense fixedpower} show that there is a significant cocoon of shock-heated material that extends down almost to the base of the jet. Indeed, the thermal energy profile begins to increase upstream of the recollimation shock, hence the on-axis shock heating cannot be the sole actor driving this thermal profile.

For a significantly higher choice of $f_{\rm J}^{\rm (thresh)} = 0.5$ (relative to our fiducial value of $10^{-3}$) the total lobe energy content of this `high' power jet drops by a factor of $5$ as more of the material in the turbulent cocoon is no longer characterised as lobe material. Crucially, however, the value of $(1/E_{\rm lobe})\;{\rm d}E_{\rm lobe,therm}(z)/{\rm d}z$ at the recollimation shock then drops by a very large factor (of $\sim18$)\footnote{Note that in the `standard' CGM case the results are much less sensitive to the adopted choice of $f_{\rm J}^{\rm (thresh)}$ as expected, given the much less developed turbulent cocoon.}. This implies that the increase in thermal energy distribution cannot be attributed solely to the thermalisation of material passing through this first recollimation shock. Rather, we expect it to be driven by the presence of hot, off-axis material in the cocoon that has already undergone shock-heating (potentially by multiple recollimation shocks)\footnote{One further point to note here is that identifying the peaks in Mach number in the simulation snapshots may miss any particularly strong shocks that thermalise on timescales shorter than the interval between snapshots.}. 

In the `high' and `medium' power jets, the thermal energy distribution increases significantly close to the base of the jet and then remains fairly constant from this point out to the head of the jet. This behaviour will be primarily due to the presence of the hot cocoon that extends down most of the length of the jets. In the case of the `low' power jet, however, the thermal energy distribution continues to increase as we move out towards its head. Comparing this to our estimate of the CGM thermal contribution we see that very little shock heating is required to explain this total thermal energy profile, indicating that the trends can be explained by the entrainment of CGM material that need not have undergone shock-heating.

\subsubsection{Mass distribution}
\label{Subsubsec: Mass distribution}

Fig.~\ref{fig: Jet mass pdf} shows the distribution of mass within the jet lobe, where the lobe mass has been split into contributions from disc, CGM and `pure' jet material. Visual inspection of Figs.~\ref{fig: Slices thick fixedpower}~and~\ref{fig: Slices thick dense fixedpower} clearly shows that there is a dense, cold sheath surrounding the jet channel close to the disc and, indeed, we can see that this feature in the mass profile can be attributed to entrained disc material. This cold sheath is slow which explains why there is no analogous peak close to the base of the jet in the energy profiles shown in Fig.~\ref{fig: Jet energy pdf}.

In the `standard' CGM runs, for the `high' and `medium' jet powers, the contribution from jet material remains subdominant along the entire length of the jet and as we move towards the head we see that CGM material becomes important as the turbulence in the cocoon acts to mix CGM material. The `low' power jet remains dominated by disc material for a greater proportion of its length. This is confirmed by the slices in Fig.~\ref{fig: Slices thick fixedpower} where we can see that the cold sheath surrounding the jet channel extends furthest along the length of the jet in `low' power case.

In the `dense' CGM case, as expected, the mass coming from CGM material plays a much more important role here than in the `standard' CGM case. CGM material becomes dominant closer to the base of the jet and remains so along its entire length, with the contribution from jet material never being significant at any location along the jet axis. Considering the `low' power jet in particular, there is a broad peak in the mass distribution towards the head of the jet, which can be associated with mixed CGM material after disruption of the jet beam as can be clearly seen in the slices of Fig.~\ref{fig: Slices thick dense fixedpower}.

\section{Blandford-Znajek jets: Results}
\label{sec: BZ results}
In the previous section we presented simulations of jets with a fixed power and direction and performed analysis of their launching from a circumnuclear disc and subsequent propagation through the surrounding CGM. Moving forward from this, we now examine simulations in which the jet power and direction evolve according to our full sub-grid model for Blandford-Znajek jets, as detailed in Sections~\ref{Sec: Theory}~and~\ref{Sec: Numerical}.

We have carried out a suite of simulations in which we probe the effects of our choice of model parameters and where we also experiment with different circumnuclear disc structures to assess the effects of enhanced accretion flow on the resultant jet morphology and evolution. A full study of the complete simulation suite is beyond the scope of this work and will be presented in a companion paper. It is, however, pertinent to employ the insight we have obtained thus far to make some concrete assessments of how relaxing the assumptions of fixed jet power and direction will affect the morphology and evolution of jets. In this section we, therefore, restrict our attention to three representative simulations which use our full Blandford-Znajek jet feedback model. 

In all of these simulations, the sub-grid $\alpha$-disc is initialised with mass $M_{\rm d}=10^3 \; {\rm M_\odot}$ and angular momentum direction $\mathbf{j}_{\rm d} = \mathbf{\hat{z}}$, while the initial black hole mass and spin are $M_{\rm BH} = 10^6 \; {\rm M_\odot}$ and $a=0.2$. This choice of $\alpha$-disc mass and angular momentum, however, does not guarantee that it will initially be in equilibrium with respect to torques from the black hole and inflow from the surroundings. Since the jet properties are sensitive to the accretion rate onto the black hole and the angular momentum carried by this material (see equations~(\ref{eq: jdotbh})~and~(\ref{eq: EdotJet})), we evolve the simulations for $2\; {\rm Myrs}$ before turning on the jet to allow the sub-grid disc to come into equilibrium. This reduces the impact of our initial choice of disc mass and angular momentum direction on the long term properties of the jet. Doing so introduces further a complication, however, namely that during these $2\; {\rm Myrs}$ the black hole mass and angular momentum (in particular the direction of its angular momentum) will also undergo evolution. To mitigate these problems, we reset the black hole spin magnitude and direction as well as the accretion disc angular momentum direction to their initial values immediately before the jet turns on. Note that we do not include these initial $2\; {\rm Myrs}$ in our analysis and henceforth, any reference to `time' refers to time elapsed since the jet was turned on.

In the first of the three simulations we present here, the black hole spin is initially parallel to the $z$-axis and the initial conditions correspond to our `standard' CGM density. In the second simulation the black hole spin is, again, initially parallel to the $z$-axis but here we use our `dense' CGM initial conditions. Finally, in the third simulation, we initialise the black hole spin direction to be perpendicular to the $z$-axis meaning that the jet is initially {\it launched into the circumnuclear disc} and here we use the initial conditions with the `standard' density CGM. For clarity, a full accounting of the parameters of our sub-grid model that differ between runs can be found in Table~\ref{tab: fullmodel}. 

Note that in the derivation of our expression for the Bardeen-Petterson torque in Section~\ref{Subsec: BP effect} we assumed that the angular momenta of the black hole and $\alpha$-disc were at most `slightly misaligned'. In the simulation where the jet is initially launched into the disc (which we set up intentionally so as to achieve the maximum effect in terms of torquing) this is evidently no longer the case. We will perform detailed investigations to determine the extent to which this assumption affects the torques in our companion paper, however, we do expect that this assumption does not qualitatively change our results but may act to increase the spin alignment timescale to some extent.

\begin{table}
 \caption{Properties of the simulations that use our full Blandford-Znajek jet model. The first and second columns show textual identifiers that we use throughout the results section when we refer to a specific simulation (e.g. the `vertical' jet launched into the `dense' CGM). The third column corresponds to the initial inclination angle of the jet direction (black hole spin) to the vertical while density of the CGM is listed in the fourth column.}
 \label{tab: fullmodel}
 \begin{tabular}{llccc}
  \hline
   Initial jet direction & CGM label & $\big(\mathbf{j}_{\rm BH} \cdot \hat{\mathbf{z}}\big)_0$ & CGM density  \\
  && [${}^\circ$]& [${\rm g \, cm^{-3}}$] \\
  \hline
   `vertical'&`standard' & $0$ & $10^{-27}$ \\
   `vertical' & `dense' & $0$ & $10^{-25}$ \\
   `into the disc' & `standard' & $90$ & $10^{-27}$ \\
  \hline
 \end{tabular}
\end{table}

\begin{figure*}
    \centering
    \includegraphics[]{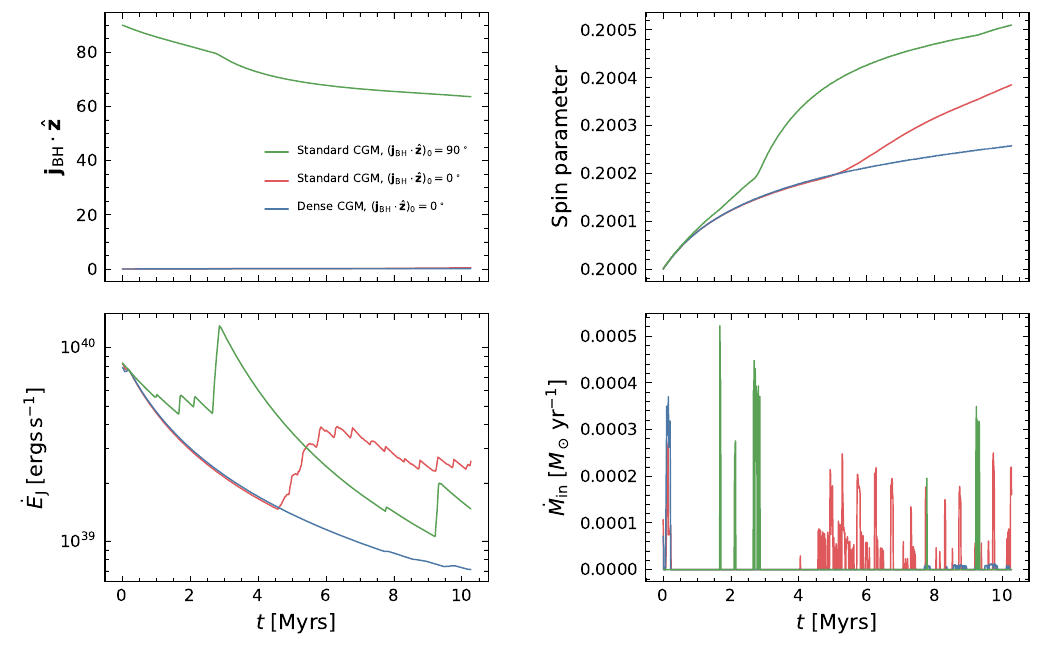}
    \caption{Time evolution of various black hole and jet properties for the three simulations that use our full Blandford-Znajek feedback model. The top-left panel shows the evolution of the angle between the jet direction and the $z$-axis. The evolution of the black hole spin parameter, $a$, is shown in the top-right panel. In the bottom-left panel we show the evolution of the jet power while the evolution of the mass inflow rate onto the $\alpha$-disc is shown in the bottom-right panel.}
    \label{Fig: Properties evolution}
\end{figure*}

\subsection{Evolution of the black hole properties}
\label{subsec: Black hole-disc evolution}
We begin our analysis by describing the evolution of the black hole-accretion disc system in all three simulations and provide discussion as to which physical processes are likely to be driving the features that we identify. We do so with the aid of Fig.~\ref{Fig: Properties evolution} which, in the top row shows the evolution of the angle between the jet direction and the $z$-axis (left-hand panel) and the evolution of the magnitude of the black hole spin (right-hand panel). In the bottom row the evolution of the jet power is shown (left-hand panel) as well as the evolution of the mass inflow rate onto the $\alpha$-disc (right-hand panel).

In all runs the initial jet power is $\sim 10^{40}\; {\rm erg\,s^{-1}}$, which is comparable to the `low' power jets\footnote{From equation~(\ref{eq: EdotJet}) it is evident that the jet power depends only on the black hole spin and mass accretion rate. Since, in all runs, the black hole spin is $0.2$ when the jet turns on, any difference in the initial jet power comes about due to variations in the equilibrium mass reached by the $\alpha$-disc during the initial relaxation period.} which were presented in Section~\ref{sec: fixed}. Before the launching of the jet there is sustained mass inflow onto the $\alpha$-disc in all runs\footnote{Note that evolution before the jet turns on is not shown in Fig.~\ref{Fig: Properties evolution}.}. Upon jet launching, however, the inflow is very quickly terminated but then resumes at later times, albeit very intermittently and with varying strengths. Considering the two `vertical' jet runs, we see that the inflows in the `dense' CGM case resume $\sim4\; {\rm Myrs}$ later and are less significant than those found in the `standard' CGM case. We examined radial velocity maps of the inner regions in these two runs and found that sustained bulk inflows are less common in the `dense' CGM case due to the higher levels of jet driven turbulence that prevent the formation of any significant coherent inflow.

Even without external inflow, the black hole continues to accrete from the $\alpha$-disc. With no mass flowing onto the $\alpha$-disc to replenish the material accreted by the black hole or entrained by the jet, the mass of the $\alpha$-disc drops. This, in turn, leads to a decrease in the accretion rate onto the black hole which, ultimately, causes the jet power to diminish, as is seen in Fig.~\ref{Fig: Properties evolution}. In all runs we observe that at later times mass is able to flow onto the system again, however, this inflow is now very sporadic and bursty. In the two runs which have a `standard' CGM density, resumption of inflow occurs after $\sim 1.7\; {\rm Myrs}$ and $\sim 4.0\; {\rm Myrs}$ for the case where the jet is launched into the disc and the case where it is launched perpendicular to the disc, respectively. At these times we observe a significant increase in the power of the relevant jet, as the black hole accretion rate increases. It is interesting to note that even in our setup (where the secularly-driven inflow occurs at a moderate rate) the jet power need not cease entirely in order for the inflow to resume. The jet is essentially `on' all the time, albeit with its power varying by approximately one order of magnitude. This illustrates how the jet power is able to naturally readjust to the properties of the gas in its immediate vicinity and, in turn, how gas properties are shaped by the jet action (e.g. via shock heating and entrainment). This suggests that jet duty cycles may not be driven solely by large-scale inflows (or lack thereof). 
\begin{figure*}
    \centering
    \includegraphics[]{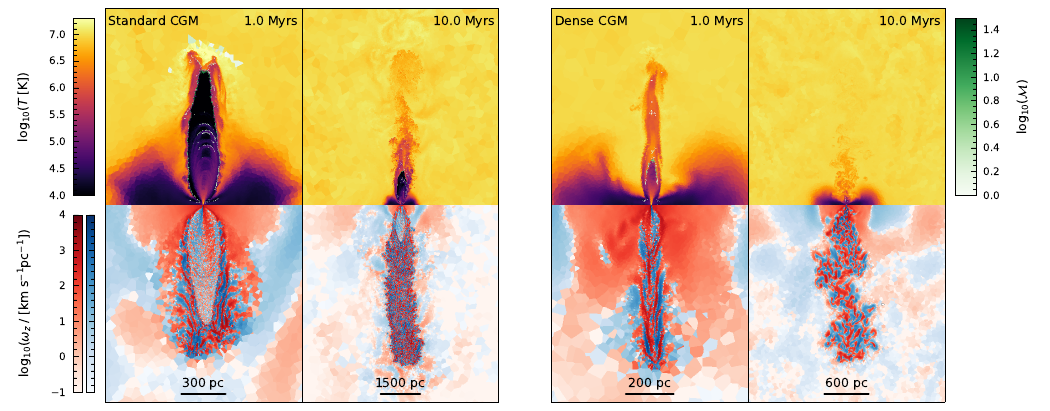}
    \caption{Slices in the $x$-$z$ plane (centred on the black hole) for the two `vertical' jets that evolve according to our full Blandford-Znajek feedback model. The two panels on the left correspond to the run in the `standard' CGM and the two on the right correspond to the run in the `dense' CGM. For each run, the first panel is created at an `early' time ($1\; {\rm Myr}$) and the second at a `late' time ($10\; {\rm Myr}$). The bottom half of each panel shows the $z$ component of the vorticity field with the colourscale corresponding to the magnitude of this field and the red and blue colours indicating oppositely directed vectors. The top half of each panel shows the temperature and onto this we overlaid the locations of any shocks (where they exist) with the green colourscale indicating the strength of the shock.}
    \label{Fig: Fullmodel vertical}
\end{figure*}

Examining the evolution of spin direction, it is evident that the direction of the jets whose axes are initialised to be vertical remains approximately unchanged throughout. This is to be expected as the large-scale gas angular momentum is well ordered and so it is unlikely that there is significant accretion of gas that has considerably misaligned angular momenta. In the run where the jet is initially directed into the disc, however, we see that the black hole spin (and thus the jet direction) is torqued towards the vertical as the Bardeen-Petterson effect acts to align it with the total angular momentum of the black hole--$\alpha$-disc system. The amount of torquing is significant, with spin direction being changed by $\sim 30^\circ$ over the course of $10\; {\rm Myrs}$. One further point to note here is there is a distinct knee in the evolution of the spin direction for this misaligned jet which is coincident with a strong burst of inflow onto the system. This highlights the fact that, as expected, the rate of alignment of the black hole angular momentum is also sensitive to the mass flows in the vicinity of the black hole.

We now turn to look at the evolution of the black hole spin magnitude and, whilst in all runs it is evident that the spin increases monotonically throughout, the rate of increase is low. This implies that there is an interesting interplay between the spin-up torques from accretion of gas from the corotating $\alpha$-disc and the spin-down torques from the launching of the Blandford-Znajek jet (we leave detailed characterisation of the relative importance of these two processes to future work). Furthermore, by construction, our simulated system can sustain only very moderate inflows onto the $\alpha$-disc as it is largely in equilibrium (and undergoes only very modest secular evolution). In more perturbed systems (such as those undergoing mergers or where large-scale cosmic inflows or cooling flows exist) we would expect that the black hole accretion rate would be higher which would then lead to more significant levels of black hole spin-up and higher jet powers.

In this section we have focused on the evolution of the black hole and jet properties but it is also instructive to understand how the $\alpha$-disc co-evolves with the black hole. Whilst we save comprehensive analysis for our future work, it is worth highlighting some key features. As mentioned above, the $\alpha$-discs are all initialised with a mass of $1000\; {\rm M_\odot}$, but during the first $2\; {\rm Myrs}$ before the jet turns on, in all three runs the discs settle into quasi-equilibrium with mass of $\sim400\; {\rm M_\odot}$. Once the jets are launched, the $\alpha$-disc masses decrease gradually as sporadic inflow events do not supply the discs with sufficient material to sustain the equilibrium disc masses that were attained when there was consistent inflow. This leads to a net decrease in the $\alpha$-disc masses of $\sim100-200\; {\rm M_\odot}$ over the course of the simulations.

When considering the net angular momentum direction of the $\alpha$-disc we find that the discs in the `vertical' jet cases remain directed along the $z$-axis, as would be expected since the black hole spins remains approximately parallel to the $z$-axis and so the disc undergoes no significant Bardeen-Petterson torque and any inflow from the circumnuclear disc is largely aligned. In the case of the jet directed into the disc, since the $\alpha$-disc also initially lies in the plane of the circumnuclear disc, it is subject to Bardeen-Petterson torques equal and opposite to those felt by the black hole. These begin to bring its angular momentum into alignment with the total angular momentum of the black hole-$\alpha$-disc system, albeit counteracted somewhat by sporadic inflow from the circumnuclear disc. It will be interesting to probe the effects of the initial $\alpha$-disc configuration in future work.

\subsection{Initially vertical jets}

We now investigate differences between the jets from Section~\ref{sec: fixed} (in which we fixed their power and axis) and those whose axes are initially vertical and evolve according to our full feedback model. For these `full' model runs, we have highlighted the fact that the spin direction does not change appreciably (see Fig.~\ref{Fig: Properties evolution}), hence any observed differences in the jet morphologies with respect to the `fixed' models are likely due to changes in the jet power alone. In terms of jet energetics the most relevant runs for comparison are the `low' power jets in the `standard' and `dense' CGM (i.e. the bottom row of Figs.~\ref{fig: Slices thick fixedpower}~and~\ref{fig: Slices thick dense fixedpower}).

\begin{figure*}
    \centering
    \includegraphics[]{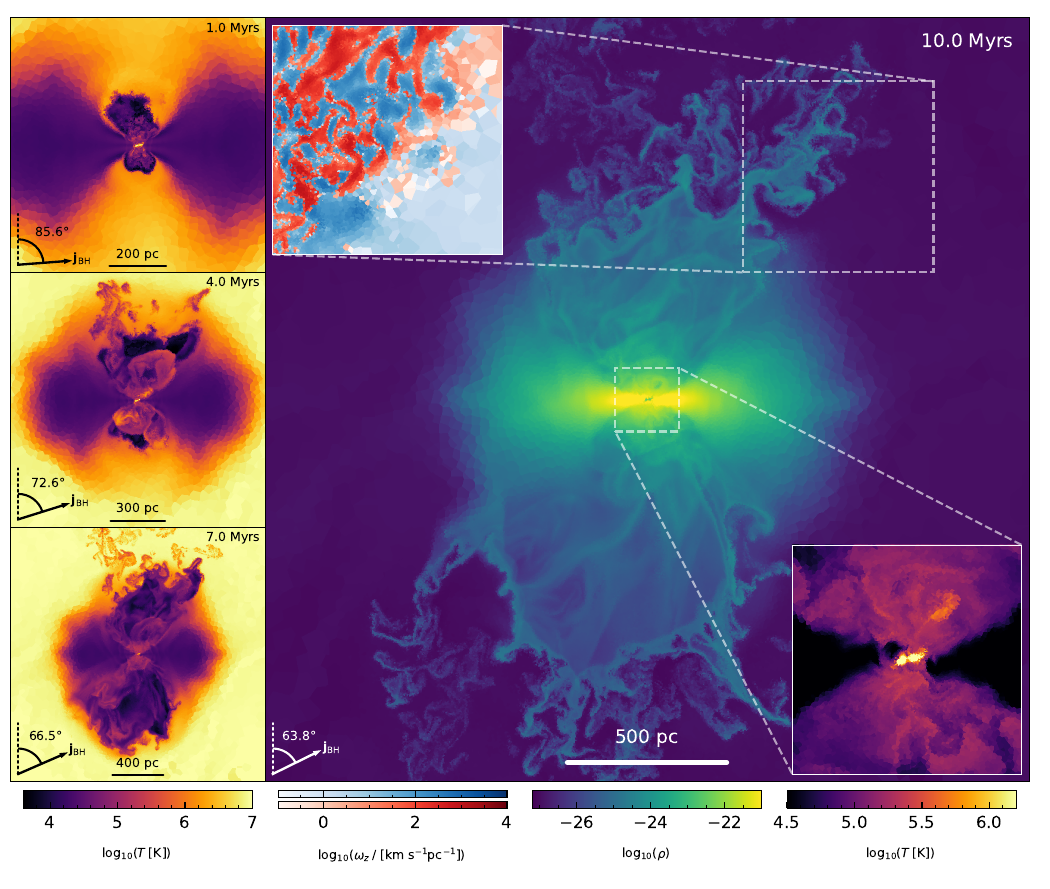}
    \caption{Slices in the $x$-$z$ plane (centred on the black hole) for the jet that is initially directed into the circumnuclear disc that evolves according to our full Blandford-Znajek feedback model. The three smaller panels on the left show slices of the temperature field, evolving in time with $3\; {\rm Myr}$ intervals from top to bottom. The main panel shows a slice of the density field after a further $3\; {\rm Myrs}$ have elapsed. The two inset plots show slices of the temperature field in the centre and the vorticity field in the outflow. In the bottom left of each panel the arrow indicates the direction of the jet (i.e. that of the black hole angular momentum) and this is labelled with the corresponding inclination angle to the vertical.}
    \label{Fig: Fullmodel horizontal}
\end{figure*}

In Fig.~\ref{Fig: Fullmodel vertical} we show slices in the $x$-$z$ plane (centred on the black hole) for the two `vertical' jets that evolve according to our `full' model. For each run we show slices at two different epochs (with the time indicated in the top right corner). The two columns on the left correspond to the run in the `standard' CGM and the two columns on the right show the run in the `dense' CGM. The bottom half of each panel shows the $z$ component of the vorticity field with the colourscale corresponding to the magnitude of this field and the red and blue colours indicate oppositely directed vectors. The top half of each panel shows the temperature and onto this we have overlaid the locations of any shocks (where they exist) with the green colourscale indicating the strength of the shock.

At $1\; {\rm Myr}$ the jet in the `standard' CGM is already undergoing recollimation and at $10\; {\rm Myrs}$ this process has indeed been completed. Comparing this to the fixed power runs, it is evident that the recollimation process occurs over shorter timescales when the jet power is allowed to decrease in accordance with the accretion rate. The fact that the CGM pressure in these two runs is the same further reinforces our previous observation (from comparisons between the `fixed' jets in `standard' and `dense' CGM cases) that the magnitude of the pressure contrast between the jet lobes and the surrounding medium is central to the recollimation process. $1\; {\rm Myr}$ after the jet is initially launched, it is still able to drive a bow shock into the CGM and there is clear evidence for internal shocks in the jet channel as shells of material that are travelling at different velocities collide. 

At this time, the typical Mach numbers along the jet axis and at the reverse shock are a factor of $\sim 2.5$  lower than found in the `low' (fixed) power jet at this time, indicating that the internal shock heating is weaker. This explains why the channel and lobes of this `full' model jet are colder than those of the `low' (fixed) power jet. After $10\; {\rm Myrs}$, the `full' model jet is no longer driving a bow shock into the CGM, however we can still distinctly see the shock that surrounds the cold jet channel. Interestingly, there is a plume of material beyond the recollimation shock which has an enhanced temperature relative to the downstream shocked jet material. This feature can be attributed to the increase in jet power that occurs due to the burst of inflow at $\sim5\; {\rm Myrs}$. This highlights the fact that jet power variability is likely to enhance levels of mixing, as new jet material interacts with the previously launched component which will ultimately affect the morphology and propagation of the jet.

Much of this discussion can also be applied to the `full' jet model in the `dense' CGM but there is one feature specific to this run that is worth highlighting. Namely, that after $10\; {\rm Myrs}$ the jet beam has been largely disrupted and mixed with the CGM material. This occurs because the jet power has dropped by an order of magnitude during the $10\; {\rm Myrs}$ since its initial launching. Whilst the late-time jet power is too low to impart any significant heat to the CGM, we see from the vorticity field that the jet is acting to stir the CGM gas which leads to enhanced levels of turbulence. This adds weight to the argument that accurate modelling of the full feedback cycle (including self-regulation) is necessary to improve our understanding of where and how jet energy is communicated to the surrounding medium and whether jet heating is localised to the centre of the galaxy or is able to regulate gas inflows from farther out. We expect that there is no `one-size-fits-all' solution to these problems, but that the dominant processes will be highly dependent on both the large and small-scale properties of the system in question. Detailed investigation of these issues will require realistic, cosmological host galaxy simulations that are evolved with self-regulated jet feedback.

\subsection{Jet into the circumnuclear disc}
\label{subsec: Jet into the disc}

Using the simulation where the jet is initially directed into the circumnuclear disc, we now briefly consider the effect on the outflow properties and jet axis evolution when the initial jet direction is not aligned with the global angular momentum of the system. We perform this analysis with the aid of Fig.~\ref{Fig: Fullmodel horizontal} in which the three panels on the left show slices of the temperature field in the $x$-$z$ plane (again, centred on the black hole), evolving in time with $3\; {\rm Myr}$ intervals from top to bottom. The main panel then shows a corresponding slice of the density field after a further $3\; {\rm Myrs}$ have elapsed with two inset plots showing the central temperature structure and the vorticity field in the outflow.

From these visualisations, it is immediately obvious that the outflow structure here is drastically different from all those considered previously in this work. Here the jet is primarily acting to blow off the surface of the disc rather than to inflate jet lobes. It is interesting to note, however, that this outflow still has a somewhat bipolar nature with the majority of the material being expelled along the $z$-axis, due to the disc pressurisation of the outflow, as is particularly clear from the temperature slices.

Inspecting the inset vorticity plot in the main panel we see that the amount of turbulence in the outflow is significantly enhanced beyond that found in the CGM. We can estimate the properties of the outflow by selecting material that satisfies our `jet lobe condition' (i.e. $f_{\rm J} > 10^{-3}$) and also has a $z$-velocity greater than $50\; {\rm kms^{-1}}$. After $10\;{\rm Myrs}$ we find that the outflowing material has mass $\sim 7\times10^3 \; {\rm M_\odot}$ which corresponds to $~0.6$ per cent of the total mass of the circumnuclear disc. Within this material we also find temperatures and densities that vary by $\sim 3$ and $\sim 6$ orders of magnitude, respectively, highlighting the multiphase nature of the outflow. The emergence of this turbulent multi-phase outflow is interesting for a multitude of reasons but two in particular stand out. Firstly, this clearly demonstrates that jets can drive galactic-scale outflows even when directed into a disc, and secondly, this hints at the possibility of dense clumps forming in the outflow which could, ultimately, lead to star formation in the outflow itself. Full investigation of the efficacy of the second scenario is beyond the scope of this study as such work would require dedicated simulations with realistic gas cooling (and star formation) in the outflow. 

We found in Section~\ref{subsec: Black hole-disc evolution} that in this simulation there is a significant burst of inflow that feeds the $\alpha$-disc around $\sim 3\; {\rm Myrs}$ after the initial launching of the jet which results in an increase in the power of the jet. In the temperature slice created at $4\; {\rm Myrs}$ in Fig.~\ref{Fig: Fullmodel horizontal} we can see the result of this increase in jet power in the form of an asymmetric outflowing shell that has temperatures enhanced above those found in the extended outflow.

In Section~\ref{subsec: Black hole-disc evolution} we also highlighted the fact that over the course of this simulation the jet axis is torqued back towards the vertical. To see this evolution more clearly we have indicated the direction of the jet (i.e. that of the black hole angular momentum) with an arrow and labelled the inclination angle in the bottom left of each panel in Fig.~\ref{Fig: Fullmodel horizontal}. At early times the jet is directed into the disc but due to the effects of sustained accretion from the $\alpha$-disc the inclination angle decreases, with distinctive asymmetric features occurring in the outflow. After $10\; {\rm Myrs}$, the jet has begun to emerge from the disc, as can clearly be seen in the inset temperature plot. 

\section{Discussion}
\label{sec: discussion}
\subsection{Comparison to previous work}
While the majority of AGN jet simulations in the context of galaxy formation typically consider larger scales and primarily focus on galaxy and cluster-scale environments, it is instructive to compare with these previous works as this allows us to highlight similarities and differences both in methodology and in physical outcomes.

Specifically, in this work we aim to improve on existing models of AGN jet feedback in galaxy-scale simulations by self-consistently predicting the jet power and direction based on the black hole properties and those of the surrounding accretion flow. We do so by explicitly following the evolution of the black hole mass and angular momentum as it accretes from a sub-grid thin (warped) $\alpha$-disc. 

Many jet studies use a simplified prescription for the jet power and direction and assume both to be fixed \citep[e.g.][]{2020BourneSijacki, 2019BambicReynolds, 2018Ehlert+, 2017BourneSijacki, 2017Weinberger+, 2002Reynolds+}. Often this choice is made out of necessity as it is not always possible to sufficiently resolve the accretion flows. Some works, however, use more physically motivated models for the jet power and assume that it is some fraction of the energy associated with the mass accreted by the black hole. An estimate for the mass accretion rate is then determined from the gas properties in the simulation and there are a variety of ways of making such an estimate in the literature; for example, \citet{2010Dubois+, 2012Dubois+} adopt the Bondi-Hoyle-like accretion rate as a measure of the hot gas accretion rate, while \citep{2016YangReynoldsB, 2017Li+, 2015Li+, 2014LiBryan} have prescriptions for cold gas accretion via a gas dropout time. Others use the instantaneous mass accretion rate at the inner boundary of the simulation domain \citep{2006VernaleoReynolds}. 

More realistic models of black hole spin evolution have begun to be incorporated into galaxy-scale simulations \citep{2019BustamanteSpringel, 2018Fiacconi+, 2014Dubois+}. More recently, \cite{2019Beckmann+} presented a model in which the jet power was determined by the black hole spin, making use of the results of GRMHD simulations to obtain a jet efficiency. There are, however, several differences with respect to our model, including their accretion prescription which is based on the Bondi-Hoyle flows and the fact that our model takes into account the back-reaction of the launching of the jet on the black hole mass.

\cite{2006VernaleoReynolds} performed idealised simulations of jets with fixed axes that are launched into cluster atmospheres and found that the jet energy was unable to couple effectively to the ICM. Some works overcome this problem by including precession \citep{2019Martizzi+, 2019Wang+, 2017RuszkowskiYang, 2014LiBryanA, 2010Falceta-Goncalves+} and others by including rapid re-orientation of the jet \citep{2018Cielo+}. These precession and reorientation prescriptions have to be put in `by hand'; in our sub-grid model, however, the jet direction is self-consistently determined by the black hole spin. Ultimately, this will allow us to assess whether the spin direction changes we observe in our simulations are sufficient to distribute the jet energy in such a way that it effectively couples to the ICM. This is particularly important as recent simulations show that effective communication of the jet energy to the ICM is possible when simulating more realistic cluster environments that include turbulence and substructures, without having to invoke jet precession \citep[e.g.][]{2020BourneSijacki, 2017BourneSijacki, 2010Morsony+}.

In order to accurately evolve the black hole properties and to determine the mass, energy and momentum injected into the jet, the mass and angular momentum fluxes in the vicinity of the black hole need to be followed at sufficiently high resolution. Due to computational constraints this requires adaptively increasing the resolution close to the black hole. Some works have successfully implemented techniques to do just this \citep{2020Angles-Alcazar+, 2019Koudmani+, 2015CurtisSijacki} and, further to this, \cite{2020BourneSijacki, 2017BourneSijacki} have successfully demonstrated the use of these techniques in combination with jet launching.

To accurately follow the inflation of the jet lobes it is essential to ensure that they are sufficiently resolved. Previous studies have found that to properly capture the energetics of the jet lobe-ICM interface and the development of instabilities, additional targeted refinement is often required \citep{2020BourneSijacki, 2018Ehlert+, 2017BourneSijacki, 2017Weinberger+}. Similarly, we also find that maintaining sufficient resolution in the jet lobes is necessary in order for the dynamics of the jet to be accurately followed. Additionally, we find that the resolution of the medium through which the jet propagates also must be sufficiently high too. Indeed, the evolution of multiphase structures and the resolution required to accurately capture its dynamics is a seemingly ubiquitous problem in astrophysics \citep{2020BennettSijacki, 2020Fielding+, 2019Ji+, 2018GronkeOh, 2017Kim+}. 

It is worth emphasising that there are many methods of injecting jets in galaxy-scale simulations. In our work, we inject mass, energy and momentum into a finite mass of material and explicitly follow the inflation of the jet lobes, as do \cite{2017BourneSijacki,2016YangReynoldsB, 2010Dubois+, 2004Omma+}. Jet models that are employed in larger scale cosmological simulations, however, necessarily mimic this lobe inflation phase and directly inject thermal energy into (off-centre) bubbles \citep{2015Schaye+, 2015Crain+, 2007Sijacki+} or impart AGN-driven kinetic winds \citep{2017WeinbergerA+}, which bypass the small scale jet thermalisation physics. Since injecting into a finite mass introduces numerical mass loading of the jet, some works implement the jet via inflow boundary conditions \citep{2012Krause+, 2010Morsony+, 2006Heinz+}.

Finally, the fact that we find such a significant change in outflow morphology when the jet is directed into the circumnuclear disc highlights the necessity of accurate modelling of this immediate environment which is typically neglected in idealised setups and not easily resolvable on cosmological scales. The quasi-bipolar, turbulent, multi-phase outflow that is launched from our circumnuclear disc when the jet is directed into it bears some resemblance to AGN-driven winds \citep{2020Costa+,2012Faucher-Giguere+, 2012ZubovasKing, 2003King} despite the fact that it is, in fact, jet-driven. Previous studies considering the interactions between inclined AGN jets and the multiphase ISM of a galactic disc found evidence for morphologically similar outflows. \cite{2018Mukherjee+} found that inclined jets launch sub-relativistic outflows along the galactic minor axis and similarly,  \cite{2018Cielo+2} found that the jet beams were deflected out of the disc plane. The fact that our simulations, and those mentioned above, find that AGN jets are able to drive outflows that closely resemble AGN winds is very intriguing and more work needs to be done to understand whether there are any observational signatures that could differentiate between these two driving mechanisms.

\subsection{Caveats}

Our sub-grid model is intended for use in galaxy-scale simulations which (generally) solve the equations of non-relativistic dynamics and in which it will not be possible to resolve the black hole horizon for the foreseeable future. This prohibits the entirely ab-initio generation of magnetised jets and so in our model we chose to specify the flux threading the black hole horizon using the results of GRMHD simulations from \cite{2012Tchekhovskoy}. As highlighted in Section~\ref{subsec: BZ}, these simulations follow the evolution of a system in which the accretion disc is in a magnetically arrested state wherein the magnetic flux on the black hole has reached saturation. Whilst this is by no means a settled issue, it is reasonable to assume, for the purposes of parameterising the magnetic flux for this model, that all black hole accretion discs are, indeed, in a magnetically arrested state \citep{2014Narayan+}. Similarly, we do not consider special relativistic effects in our jet model as the velocities in all the jets considered in this work never exceed $0.2\;c$. Any such effects will therefore have negligible impact on the propagation of our jets.

Our sub-grid disc is modelled as a thin $\alpha$-disc which allows us to follow the angular momentum flux of accreted material onto the black hole, meaning that the black hole spin can be accurately evolved. This would not be possible were we to use an accretion prescription that assumes spherical symmetry. Knowledge of the black hole spin is required for the Blandford-Znajek jet model (along with the mass accretion rate), and so our $\alpha$-disc accretion prescription is what, ultimately, allows us to couple the accretion flow to the jet properties. 

Much theoretical and observational work has gone into the determination of the properties of black hole accretion flows in systems which are capable of producing radio jets. A possible picture is that this radio mode feedback coincides with low Eddington ratios, where electrons and ions decouple into a two-temperature fluid and mass is accreted through a radiatively inefficient, thick disc\footnote{Super-Eddington accretion is also likely to lead to the formation of a thick disc due to the larger optical depths \citep{1988Abramowicz+}.} \citep{1994NarayanYi, 1976Shapiro+}. In this picture, thin accretion discs are predominantly associated with quasar mode feedback. 

It is not, however, certain whether the existence of a radio-jet necessitates the presence of a thick accretion disc. Recent optimisations of GRMHD codes have allowed simulations that resolve the thin disc structure to be performed \citep{2020Liska+,2019Liska+} and these studies show that thin accretion disc can indeed launch magnetically dominated jets. Note that some GRMHD simulations also find that the presence (or lack thereof) of radio jets is highly dependant on the initial magnetic field configuration which is what primarily determines whether a large-scale magnetic flux is able to build up on the black hole \citep{2014YuanNarayan}. Further to this, the quasar-radio dichotomy primarily stems from observations of X-ray black hole binaries \citep{2004Fender+} and it is not clear whether this observational picture necessarily extrapolates to the case of supermassive black holes, with FRII sources being an obvious counter-example \citep{2008MerloniHeinz, 2006Hardcastle+}. Future theoretical effort, in which both thin and thick discs are modelled self-consistently, will be essential in order to better explore this issue.

In order to understand the workings of our model and in some cases, due to the additional physical and computational complexity, we neglected several physical processes in our simulations. Firstly, our simulations were evolved according to a purely hydrodynamic model which does not include the effects of magnetic fields. It is expected that magnetic fields act to suppress instabilities \citep[e.g.][]{2008DursiPfrommer, 2006Lyutikov} and thus may play a significant role in determining the evolution of the jet-ICM interface. This has been highlighted in recent jet propagation studies that include magnetic field evolution \citep{2018Ehlert+,2017Weinberger+,2016English+}. Neglecting magnetic fields has allowed us to focus specifically on the workings of our model. It will, however, be important to include such effects in future work, especially as doing so could provide constraints on the magnetic flux threading the black hole horizon which could be used to improve our feedback model. 

We also do not model the dynamics of cosmic rays in our simulations. The non-thermal pressure associated with cosmic ray protons may be significant in the jet lobes and the excitation and subsequent damping of Alv\'{e}n waves could play a significant role in energy transport to the ICM \citep{2014BrunettiJones, 2007Ensslin+, 2004Brunetti+}. Recent jet studies have included the effects of cosmic ray pressure (and transport) \citep{2019Yang+, 2018Ehlert+, 2017Weinberger+, 2008Sijacki+}. Much work is also being done to develop improved cosmic ray transport theories \citep{2020Thomas+,2019ThomasPfrommer,2018JiangOh} and incorporate these into numerical schemes \citep{2020Hopkins+, 2020Thomas+, 2019Dubois+, 2017Pfrommer+}, meaning that it will be vital to incorporate cosmic ray physics moving forwards with our model.

Our simulations also do not model any physical viscosity which means that we are not able to accurately capture the rate at which sound waves are damped \citep{2020Berlok+, 2018Zweibel+, 2006SijackiSpringel}. It is, however, important to consider this effect when assessing the relative importance of the damping of sound waves as a method to communicate the jet energy to the ICM. Accurate modelling of physical viscosity in weakly collisional plasma (such as the ICM) is notoriously hard and sensitively depends on the magnetic field dynamics and topology \citep{2019Zhuravleva+, 2014MogaveroSchekochihin}. For similar reasons we have also neglected thermal conduction \citep{2017Kannan+, 2010RuszkowskiOh} which, in combination with magnetic fields, is expected to be anisotropic, favoured along the direction of the magnetic field lines and limited across the interface that separates the jet from the ICM/CGM due to magnetic draping \citep{2008DursiPfrommer}.

\section{Conclusions}
\label{sec: conclusions}
In this work we have developed a self-consistent sub-grid model for black hole accretion through a (warped) $\alpha$-disc and feedback in the form of a kinetic Blandford-Znajek jet within the moving mesh code $\textsc{arepo}$. Our simulation setup has been chosen to represent an environment analogous to that found at the centre of radio-loud Seyferts. We considered jets launched by a black hole located at the centre of a circumnuclear disc. This circumnuclear disc is itself embedded within a hot CGM and we varied the density of this medium to consider the effects of increased/decreased pressure on the jet propagation. We first performed simplified simulations in which we fixed the jet power and direction, allowing us to more straightforwardly compare our results with analytic predictions and identify the dominant physical processes at play. Finally, we examined three exemplary simulation cases in which we employed our full Blandford-Znajek model, with jets initially perpendicular to or directed into the plane of the circumnuclear disc. 

Our main results regarding the jets with fixed power and direction are:
\begin{itemize}
\item Jets that are significantly overpressurised with respect to the surrounding CGM initially terminate in a strong reverse shock. At later times they tend to recollimate and this process occurs on a faster time-scale if the pressure contrast between the jet and the CGM is lower. Recollimation excites instabilities at the boundary of the jet channel, driving material off-axis and feeding the surrounding cocoon of diffuse, shock-heated gas. At higher jet powers, a series of recollimation shocks form along the jet axis as the jet beam is able to retain significant ram pressure component. At lower powers, however, the jet beam is disrupted, and in this case the prevailing action of the jet is to stir the CGM and drive turbulence in the cocoon. 

\item The majority of jet lobe energy is initially in kinetic form, however, over the course of the simulations we find that a significant fraction of energy can thermalise due to shock heating and mixing (primarily with CGM material). Shock heating is more effective at higher jet powers (and for more overpressurised jets), whereas at lower powers the thermal energy associated with mildly shocked (or even unshocked) entrained CGM material is sufficient to explain the total thermal energy of the jet lobes. Once a large thermal component builds in the jet lobes this energy can be communicated efficiently with the surrounding medium.

\item  The initial evolution of largely overpressurised jets is fairly well described by the standard analytic models \citep[e.g.][]{1989BegelmanCioffi}. Once recollimation occurs, however, our simulations indicate that a key assumption inherent in the analytic models is violated: namely, the momentum flux at the base of the jet is {\it not} the same as at the head of the jet and this momentum flux {\it does} evolve with time.
    
\item As jets initially break out from the circumnuclear disc they entrain cold disc material and in some cases not-insignificant quantities of this disc material can be present along the entire length of the jet. While energetically this entrained disc material is largely irrelevant, it dominates the total jet lobe mass, until (in some cases) sufficient CGM material becomes mixed in, which occurs largely as a consequence of the excitation of instabilities by the recollimation process.     
\end{itemize}

Now considering the jets which evolve according to our full Blandford-Znajek model, the main results are as follows:
\begin{itemize}
\item When the power and direction of the AGN jets are determined self-consistently, the jet morphology and evolution differs significantly from those of jets in which the power and direction are fixed, even when the jet direction does not evolve significantly. These differences are substantial enough that the evolution of these self-regulated jets cannot be predicted using `fixed power and direction' jet models. Initially, as the jet emerges from the circumnuclear disc it cuts off gas supply to the black hole and the jet power consequently drops significantly as the $\alpha$-disc is drained. This leads to slower jet velocities, weaker shocks and colder material in the jet channel. 

\item The jet power is subject to self-regulation: as a consequence of the secular evolution of the circumnuclear disc, inflowing gas is eventually able to reach the $\alpha$-disc again causing the jet power to increase significantly. Hereafter, the inflow is intermittent and bursty, but can sustain a continuous jet (albeit with varying power) with the supermassive black hole gradually spinning up. 
   
\item For the jet initially launched directly into the circumnuclear disc (as may be the case if the black hole spin was initially misaligned with respect to the global angular momentum) the resulting outflow bears little resemblance to the outcomes previously considered. The jet drives a turbulent, multi-phase, quasi-bipolar outflow as it blows-off the outer layers of the circumnuclear disc. Accretion onto the $\alpha$-disc persists and the jet is torqued efficiently out of the circumnuclear disc. As it emerges, its collimated outflow propagates into the previously generated multi-phase wind. 
\end{itemize}

Our modelling of Blandford-Znajek jets opens up many new avenues through which the role of AGN feedback in galaxy formation can be explored. We have shown here that self-consistent jet power evolution is a natural consequence when black hole accretion and spin evolution prescriptions are coupled to the jet feedback. Ultimately, this leads to the energetics and direction of the jet being determined by, but also acting to shape the properties of the wider environment. In this work we have explored a scenario in which the central region of the galaxy is largely in equilibrium, with gas inflows being driven by modest secular evolution. It will be particularly interesting to apply our model to more dynamically evolving environments such as major mergers and cooling flow clusters. This will improve our understanding of the extent to which self-regulated AGN jet feedback influences the large-scale properties of these systems and help us determine the dominant processes by which it does so. We have identified three largely distinct channels through which these jets shape their surroundings: (i) direct heating via shocks, (ii) mixing and turbulence, and (iii) multi-phase, galactic-scale outflows in which the innermost regions of the galactic disc are obliterated and drawn up into the outflow. A natural next step will be to apply this model in the context of full cosmological simulations with the ultimate aim of providing stronger constraints on the relative importance of these interaction channels and improving our understanding of the role these jets play in shaping the properties of galaxy populations as a whole.

\section*{Acknowledgements}
We would like to thank Martin Haehnelt and Chris Reynolds for helpful comments on the manuscript. RYT, MAB and DS acknowledge the support from the ERC starting  grant 638707 `Black holes and their host galaxies: co-evolution  across  cosmic time' and the STFC. This work was performed using the Cambridge Service for Data Driven Discovery (CSD3), part of which is operated by the University of Cambridge Research Computing on behalf of the STFC DiRAC HPC Facility (www.dirac.ac.uk). The DiRAC component of CSD3 was funded by BEIS capital funding via STFC capital grants ST/P002307/1 and ST/R002452/1 and STFC operations grant ST/R00689X/1. DiRAC is part of the National e-Infrastructure. This work used the DiRAC@Durham facility managed by the Institute for Computational Cosmology on behalf of the STFC DiRAC HPC Facility (www.dirac.ac.uk). The equipment was funded by BEIS capital funding via STFC capital grants ST/P002293/1 and ST/R002371/1, Durham University and STFC operations grant ST/R000832/1. DiRAC is part of the National e-Infrastructure.

\section*{Data availability}
The data underlying this article will be shared upon request to the corresponding author.




\bibliographystyle{mnras}
\bibliography{references}



\appendix
\section{Spin dependent parameters}
\label{App: Spin dep params}
The spin dependent ISCO radius is given by
\begin{equation}
    R_{\rm ISCO}(a) = \Lambda(a)\frac{G\, M_{\rm BH}}{c^2}\, ,
\end{equation}
where
\begin{equation}
    \Lambda(a) = 3 + Z_2(a) \mp \sqrt{(3-Z_1(a))(3+Z_1(a)+2Z_2(a))} \, .
\end{equation}
In this expression the upper and lower subtraction and  addition operators differentiate between prograde and retrograde motion cases, correspondingly. $Z_1(a)$ and $Z_2(a)$ are given by 
\begin{equation}
    Z_1(a) = 1 + (1-a^2)^{1/3}\big[(1+a)^{1/3}+ (1-a)^{1/3}\big] \, ,
\end{equation}
and
\begin{equation}
    Z_2(a) = \sqrt{3a^2 + Z_1^2(a)}.
\end{equation}
The spin dependent radiative efficiency is then
\begin{equation}
    \epsilon_{\rm r}(a) = 1-\sqrt{1 - \frac{2}{3\Lambda(a)}} \, .
\end{equation}
Finally, the specific angular momentum at the ISCO is
\begin{equation}
    L_{\rm ISCO}(a) = \pm \frac{G M_{\rm BH}}{c\Lambda(a)}\frac{\Lambda(a)^2 \mp 2a\sqrt{\Lambda(a)}+a^2}{\sqrt{\Lambda(a) - 3 \pm 2 a/ \sqrt{\Lambda(a)}}} \, ,
\end{equation}
where, again, the upper and lower addition and subtraction operators differentiate between prograde and retrograde motion.

\section{The Blandford-Znajek energy and angular momentum fluxes}
\label{App: Derivation}
Throughout this section we use comma and semicolon notation for partial and covariant derivatives. For clarity, we use units such that $G=M_{\rm BH}=c=1$ in the initial calculations but reintroduce physical units at the end. Throughout the derivation, we work in Kerr-Schild (horizon-penetrating) coordinates. Further, we assume that the fields are stationary, axisymmetric, symmetric about $\theta = \pi/2$ and that the black hole magnetosphere is force-free.

The quantities we wish to determine are the radial fluxes of energy and angular momentum across the black hole horizon, as measured by a stationary observer at infinity
\begin{align}
    \label{eq: e flux}
   &\dot{E}_{\rm BZ} = -\int_{r=r_{\rm H}} T^r{}_t \sqrt{-g}\;\D\theta\D\phi \, \\
   \label{eq: j flux}
   &\dot{J}_{\rm BZ} = -\int_{r=r_{\rm H}} T^r{}_\phi \sqrt{-g}\;\D\theta\D\phi \, ,
\end{align}
where $T^{\mu\nu}$ is the stress-energy tensor, $r_{\rm H} = r_{\rm g}(1 + \sqrt{1 - a^2})$ is the radial coordinate of the black hole horizon and $g$ is the determinant of $g^{\mu\nu}$, the metric tensor. 

Under the force-free condition $T^{\mu\nu}$ reduces to the electromagnetic stress-energy tensor, $T_{\rm EM}^{\mu\nu}$, which is given by
\begin{equation}
    T_{\rm EM}^{\mu\nu} = F^{\mu\gamma}F^{\nu}{}_{\gamma} - \frac{1}{4}g^{\mu\nu}F^{\alpha\beta}F_{\alpha\beta}\; ,
\end{equation}
where $F_{\mu\nu}$ is the Faraday tensor\footnote{We have absorbed a factor of $\frac{1}{\sqrt{4\pi}}$ into the definition of the Faraday tensor.} which, in terms of the electromagnetic four-potential, $A^\mu$, is
\begin{equation}
    F_{\mu\nu} = A_{\nu,\mu} - A_{\mu,\nu} \; ,
\end{equation}
Henceforth we will drop the subscript `EM' in the electromagnetic stress-energy tensor.

The assumption of ideal MHD gives the constraint
\begin{equation}
\label{eq: Ideal}
    {}^*F^{\mu\nu}F_{\mu\nu} = 0 \, ,
\end{equation}
where ${}^*F^{\mu\nu}$ is the dual of the Faraday tensor. It then follows that
\begin{equation}
    A_{\phi,\theta}A_{t,r} = A_{t,\theta}A_{\phi,r} \, ,
\end{equation}
and $A_t$ is a function of $A_\phi$ and therefore constant on magnetic surfaces (surfaces defined by $A_\phi = \text{const}$). Thus, we define
\begin{equation}
\label{eq: Ferraro}
    -\omega(r,\theta) \equiv \frac{A_{t,\theta}}{A_{\phi,\theta}}=\frac{A_{t,r}}{A_{\phi,r}} \, ,
\end{equation}
such that $\omega(A_\phi)$ is a function that is also constant on magnetic surfaces.

Imposing stationarity and axisymmetry then gives $F_{\phi t} = F_{t\phi} = 0$. The remaining non-zero components of the Faraday tensor can then be written in terms of the three variables $A_\phi$, $\omega$ and $B^\phi$ as
\begin{align}
    &F_{r\phi} = -F_{\phi r} = A_{\phi,r} \, , \nonumber \\
    &F_{\theta\phi} = -F_{\phi\theta} = A_{\phi,\theta}\, , \nonumber\\
    &F_{tr} = -F_{rt} = \omega A_{\phi,r}\, , \nonumber \\
    &F_{t\theta} = -F_{\theta t} = \omega A_{\phi,\theta} \, , \nonumber\\
    &F_{r\theta} = -F_{\theta r} = \sqrt{-g} B^\phi \, ,
\end{align}
and we also identify
\begin{align}
    &B^r = \frac{1}{\sqrt{-g}}A_{\phi,\theta}\, ,\nonumber \\
    &B^\theta = -\frac{1}{\sqrt{-g}}A_{\phi,r} \, .
\end{align}
The relevant components of the stress-energy tensor, evaluated on the black hole horizon, are then 
\begin{align}
    &T^r{}_t =-2\bigg(\frac{A_{\phi,\theta}}{\sqrt{-g}}\bigg)^2r_{\rm H}\omega (\omega - \Omega_{\rm H})\sin^2\theta\, , \\
    &T^r{}_\phi =\frac{1}{\omega} \; T^r{}_t\; .
\end{align}
The evaluation of equations~(\ref{eq: e flux})~and~(\ref{eq: j flux}) therefore require the determination of the fields $A_\phi$, $\omega$ and $B^\phi$. This is done by solving the equation of motion
\begin{equation}
    T^{\mu\nu}_{\rm EM \, ;\nu} = 0 \; .
\end{equation}
The resulting equations are highly non-linear and we hence proceed, as in \cite{1977BlandfordZnajek}, by perturbing the known split-monopole solution for the non-spinning case (Schwarzschild solution) and take quantities to some order of in $a$.

For this $a=0$ case, the equation of motion reduces to
\begin{equation}
    \mathcal{L}A_\phi^{(0)} = 0 \, ,
\end{equation}
where we use the superscript $(0)$ to indicate solutions to this non-spinning case and $\mathcal{L}$ is the linear operator
\begin{equation}
    \mathcal{L}\equiv \frac{1}{\sin\theta}\pd{r}\bigg[\bigg(1-\frac{2}{r}\bigg)\pd{r}\bigg] + \frac{1}{r^2}\pd{\theta}\bigg[\frac{1}{\sin\theta}\pd{\theta}\bigg] \, .
\end{equation}
The solutions for a split-monopole field geometry are
\begin{align}
    &A_\phi^{(0)}(r,\theta) = -C \cos\theta\, , \\
    &\omega^{(0)}(r,\theta) = 0\, ,\\
    &B^{\phi (0)}(r,\theta) = 0 \, ,
\end{align}
where $C$ is a constant.

Perturbing this to the next highest order in $a$, the fields take the form
\begin{align}
    &A_\phi(r,\theta) = -C \cos\theta + A_\phi^{(2)}(r,\theta)a^2 + \mathcal{O}(a^4)\, ,\\
    &\omega(r,\theta) = \omega^{(1)}(r,\theta) a + \mathcal{O}(a^3)\, ,\\
    &B^{\phi}(r,\theta) = B^{\phi (1)}(r,\theta)a + \mathcal{O}(a^3) \, .
\end{align}
$A_\phi$ is even in $a$  since the radial magnetic field geometry should not depend on the direction of the rotation. Symmetry arguments further constrain $\omega$ and $B^\phi$ to be odd functions of $a$. \\[5pt]
To $\mathcal{O}(a^2)$ the equation of motion is
\begin{equation}
    \label{eq: gs2}
    \mathcal{L}A_\phi^{(2)} = S(r,\theta) \, ,
\end{equation}
where $S(r,\theta)$ is a source function that depends on $\omega^{(1)}$ and $B^{\phi (1)}$. \cite{1977BlandfordZnajek} proceed by requiring that the fields are finite on the horizon and then match this solution to the flat-space solution at infinity \citep{1973Michel}. 

Imposing these conditions specifies $\omega^{(1)}$ and $B^{\phi (1)}$, and $A_\phi^{(2)}$ can thus be found by inverting equation~(\ref{eq: gs2}) using the known Green's function for the linear operator. Altogether this gives
\begin{align}
    \label{eq: A1}
    &A_\phi(r,\theta) = -C \cos\theta + C f(r)\cos\theta\sin^2\theta\, a^2 + \mathcal{O}(a^4)\, ,\\
    \label{eq: w1}
    &\omega(r,\theta) = \frac{1}{8}a + \mathcal{O}(a^3)\, ,\\
    \label{eq: B1}
    &B^{\phi}(r,\theta) = -\frac{C}{8r^2}\bigg(1+\frac{4}{r}\bigg)a +\mathcal{O}(a^3) \, ,
\end{align}
where
\begin{align}
    f(r) =& \Bigg[Li_2\bigg(\frac{2}{r}\bigg) - \ln\bigg(1-\frac{2}{r}\bigg)\ln\bigg(\frac{r}{2}\bigg)\Bigg]\frac{r^2(2r-3)}{8}\\
    &+ \frac{1+3r-6r^2}{12}\ln\bigg(\frac{r}{2}\bigg) + \frac{11}{72}+ \frac{1}{3r} + \frac{r}{2} - \frac{r^2}{2} \, ,
\end{align}
and $Li_2(x)$ is the second polylogarithm function
\begin{equation}
    Li_2(x) = -\int_0^1\frac{\ln(1-tx)}{t}\D t \, .
\end{equation}
Since the flux integrals only require knowledge of the horizon fields, it is important to note that the value of $f$ evaluated at $r = r_{\rm H}$ 
\begin{equation}
    f(r_{\rm H}) = \frac{6\pi^2-49}{72}\, .
\end{equation}
The first correction to the outward fluxes of electromagnetic energy and angular momentum are then found by inserting these results into equations~(\ref{eq: e flux})~and~(\ref{eq: j flux}), and evaluating the resulting integrals to the lowest order in $a$
\begin{align}
    \dot{E}_{\rm BZ} &= 8\pi\int_0^{\frac{\pi}{2}} \frac{1}{\sqrt{-g}}A_{\phi,\theta}^2 \omega(\Omega_{\rm H}-\omega)r_{\rm H}\sin^2\theta \, \D\theta \nonumber\\
    &= \frac{\pi}{24}a^2C^2+ \mathcal{O}(a^4) \, ,
\end{align}
and 
\begin{align}
   \dot{J}_{\rm BZ} &= 8\pi\int_0^{\frac{\pi}{2}} \frac{1}{\sqrt{-g}}A_{\phi,\theta}^2 (\Omega_{H}-\omega)\sin^2\theta \, r_{\rm H} \D\theta \nonumber\\
    &= \frac{\pi}{3}aC^2+ \mathcal{O}(a^3) \, .
\end{align}

For high spin values, a more accurate expression for these fluxes would be found by expanding the fields to higher order in $a$. Doing so requires boundary conditions to be imposed in order to determine the higher order expansions of the fields and it not obvious how to do so \citep{2008TanabeNagataki}. We can, however, find the expansion of $\dot{E}_{\rm BZ}$ to $\mathcal{O}(a^4)$ since the undetermined quantities in the higher order expansion of the fields do not come into the energy flux at this order. We find
\begin{align}
    \dot{E}_{\rm BZ} &= 8\pi\int_0^{\frac{\pi}{2}} \frac{1}{\sqrt{-g}}A_{\phi,\theta}^2 \omega(\Omega_{H}-\omega)\sin^2\theta \, r_{\rm H} \D\theta \nonumber\\
    &= \frac{\pi}{24}C^2a^2 + \bigg(\frac{56-3\pi^2}{1080}\bigg)\pi a^4C^2 + \mathcal{O}(a^6) \, .
\end{align}
These undetermined fields do, however, come into the second correction of the angular momentum flux and so, without making further assumptions, we cannot proceed further.

For the purposes of our sub-grid model, it is useful to express these fluxes in terms of $\Omega_{\rm H}$ \citep{2010Tchekhovskoy+} and we do so using the expansion
\begin{equation}
    a =  4\Omega_{\rm H} - 16\Omega_{\rm H}^3 +64\Omega_{\rm H}^5 + o(\Omega_{\rm H}^7) \, .
\end{equation}
Identifying $2\pi C =\Phi_{\rm BH}$ and reintroducing the factor of $1/\sqrt{4\pi}$ that was absorbed into the Faraday tensor gives
\begin{align}
&\dot{E}_{\rm BZ} =\frac{1}{24\pi^2} \Phi_{\rm BH}^2\Omega_{\rm H}^2 + \frac{(67 -6\pi^2)}{135\pi^2} \Phi_{\rm BH}^2\Omega_{\rm H}^4 + \mathcal{O}(\Omega_{\rm H}^6)\, , \\ 
&\dot{J}_{\rm BZ} = \frac{1}{12\pi^2}\Phi_{\rm BH}^2\Omega_{\rm H} + \mathcal{O}(\Omega_{\rm H}^3) \, .
\end{align}

Thus far we have assumed a split-monopole configuration. A general magnetic field configuration alters the fluxes only via the introduction of a factor $\kappa$ which in the case of a split-monopole configuration is $\kappa = 1/(6\pi)$. In terms of $\kappa$ the fluxes are then
\begin{align}
&\dot{E}_{\rm BZ} =\frac{\kappa}{4\pi}\bigg( \Phi_{\rm BH}^2\Omega_{\rm H}^2 + \frac{24(67 -6\pi^2)}{135} \Phi_{\rm BH}^2\Omega_{\rm H}^4\bigg) + \mathcal{O}(\Omega_{\rm H}^6)\, , \\
&\dot{J}_{\rm BZ} = \frac{\kappa}{2\pi}\Phi_{\rm BH}^2\Omega_{\rm H} + \mathcal{O}(\Omega_{\rm H}^3) \, .
\end{align}
Finally, we dimensionalise these equations and write them in terms of the dimensionless magnetic flux
\begin{align}
    \dot{E}_{\rm BZ} =\frac{\kappa}{4\pi}\phi^2_{\rm BH} \dot{M}_{\rm BH,0} c^2\Bigg[&\bigg(\frac{a}{2(1+\sqrt{1-a^2})}\bigg)^2 \nonumber \\
    +\frac{24(67 -6\pi^2)}{135} &\bigg(\frac{a}{2(1+\sqrt{1-a^2})}\bigg)^4\Bigg] + \mathcal{O}(\Omega_{\rm H}^6)\, , 
\end{align}
\begin{equation}
    \dot{J}_{\rm BZ} = \frac{\kappa}{2\pi}\phi_{\rm BH}^2\bigg(\frac{a}{2(1+\sqrt{1-a^2})}\bigg)\frac{G M_{\rm BH}}{c}\dot{M}_{\rm BH,0} + \mathcal{O}(\Omega_{\rm H}^3) \, ,  
\end{equation}
where we have used the fact that 
\begin{equation}
    \Omega_{\rm H} = \frac{ac}{2r_{\rm g}(1+\sqrt{1-a^2})} \, .
\end{equation}


\bsp	
\label{lastpage}
\end{document}